\documentclass[preprint]{aastex}
\usepackage{epsfig}
\usepackage{natbib}
\shortauthors{Frinchaboy et al.}
\shorttitle{Kinematics of the Sagittarius dSph }

\begin{document}

\title{A 2MASS All-Sky View of the Sagittarius Dwarf Galaxy: VII. \\ 
Kinematics of the Main Body of the Sagittarius dSph}

\author{Peter M. Frinchaboy\altaffilmark{1,2,3,4}, Steven R. Majewski\altaffilmark{5}, 
Ricardo R. Mu\~noz\altaffilmark{6,7}, David R. Law\altaffilmark{8}, \\
Ewa L. {\L}okas\altaffilmark{9},
William E. Kunkel\altaffilmark{10},  Richard J. Patterson\altaffilmark{5},
and Kathryn V. Johnston\altaffilmark{11}}

\email{ p.frinchaboy@tcu.edu, srm4n@vigrinia.edu,
  rmunoz@das.uchile.cl, drlaw@di.utoronto.ca, lokas@camk.edu.pl, kunkel@lcoeps1@lco.cl,
  rjp0i@vigrinia.edu, kvj@astro.columbia.edu 
}

\altaffiltext{1}{Department of Physics \& Astronomy, Texas Christian University,
TCU Box 298840, Fort Worth, TX 76129}

\altaffiltext{2}{National Science Foundation Astronomy and Astrophysics Postdoctoral Fellow, 
University of Wisconsin--Madison, Department of Astronomy, 5534 Sterling Hall, 475 N.\ Charter Street, Madison, WI 53706. }

\altaffiltext{3}{Visiting Astronomer, Cerro Tololo Inter-American Observatory, National 
Optical Astronomy Observatory, which is operated by the Association of Universities for 
Research in Astronomy, Inc. (AURA) under cooperative agreement with the National Science Foundation. }

\altaffiltext{4}{Any opinions, findings, and conclusions or recommendations 
expressed in this material are those of the author(s) and 
do not necessarily reflect the views of the National Science 
Foundation.}

\altaffiltext{5}{Department of Astronomy, University of Virginia, 
P.O. Box 400325, Charlottesville, VA 22904-4325, USA }

\altaffiltext{6}{Astronomy Department, Yale University,
P.O. Box 208101, New Haven, CT 06520-8101 USA}

\altaffiltext{7}{Departamento de Astronom\'ia, Universidad de Chile,
Casilla 36-D, Santiago, Chile (current address)}

\altaffiltext{8}{Dunlap Institute for Astronomy \& Astrophysics, University of Toronto, 50 St. George Street, Toronto M5S 3H4, Ontario, Canada}

\altaffiltext{9}{Nicolaus Copernicus Astronomical Center, Bartycka 18, 00-716 Warsaw, Poland }

\altaffiltext{10}{Las Campanas Observatory, Casilla 601, La Serena, Chile}

\altaffiltext{11}{Department of Astronomy, Columbia University, New York, NY 10027, USA}

\begin{abstract}
We have assembled a large-area spectroscopic survey of giant stars in the Sagittarius (Sgr) dwarf galaxy
core.  Using medium resolution ($R \sim 15,000$), multifiber spectroscopy we have
measured velocities 
of these stars, which extend up to 12
degrees from the galaxy's center (3.7 core radii or 0.4 times the King
limiting radius). 
From these high quality spectra we identify
1310 Sgr members out of 2296 stars surveyed distributed across 24 different fields across the Sgr core.  
Additional slit spectra were obtained of stars bridging from the Sgr core to its trailing tail.
Our systematic, large area sample 
shows no evidence for 
significant rotation, 
a result at odds with
the $\sim 20$ km
s$^{-1}$ rotation required as an explanation for the
bifurcation seen in the Sgr tidal stream; the observed small ($\le 4$ km s$^{-1}$)
velocity trend along primarily the major axis is consistent with models of the projected motion of 
an extended body on the sky with no need for intrinsic rotation.
The Sgr core is found to have
a flat velocity dispersion  (except for a kinematically colder center point)
across its surveyed extent and into its tidal tails, a property
that matches the velocity dispersion profiles measured for other Milky Way dwarf
spheroidal (dSph) galaxies.  We comment on the possible significance of this observed kinematical
similarity for the dynamical state of the other classical Milky Way dSphs in light of the fact that Sgr is clearly a 
strongly tidally disrupted system.

\end{abstract}

\section{ INTRODUCTION }

The Sagittarius (Sgr) dwarf galaxy is a satellite of the Milky Way (MW) with an 
apparent
dwarf spheroidal (dSph) morphology that is, however, distinct in at least two
ways.  First, the relative proximity of Sgr to the Sun ($d \approx 24$ kpc)
makes its brightest stars
quite accessible to study with even relatively modest aperture telescopes.  Thus, 
Sgr is open to much more straightforward and extensive examination than its 
much more distant, counterpart MW satellites 
making Sgr, in principle, 
a highly approachable, local laboratory for dSph galaxies.  
Second, Sgr represents the most certain and vivid example of a satellite enduring
tidal disruption by the MW, with the highest known surface brightness tidal tails 
among MW dSph satellites.  Indeed, not only does no other MW dSph have anywhere near as 
obvious tidal debris, whether other dSphs are showing any evidence of visible mass loss 
is still debated for most MW satellites.
For this reason, Sgr is often relegated to a ``special case'' category among MW satellites
and investigators of the properties of ``classical dSphs'' often avoid Sgr because of an
uncertainty and wariness of how it fits into the larger dSph context.
Obviously prudence dictates care in such matters, but the
potential advantages of a 
much more accessible test laboratory for dSph
dynamics makes it worth understanding whether the second of the above two Sgr properties
(tidal disruption) truly nullifies the usefulness
of the first (proximity) in terms of adopting Sgr as a valid prototype for other dSph satellites.

An early analysis of the internal dynamics of Sgr was provided by
Ibata et al. (1997).   
This impressively assembled early sample of radial velocities
numbered some 410 stars using the CTIO/ARGUS and AAT/AUTOFIB
spectrographs, and it remains the only fully published wide-spread
detailed analysis of the dynamical properties of the Sgr galaxy main body.  
While the area covered provides some strength to look at the larger
Sgr main body, the limitation of most fields being along the main axis
limits further analysis, as does the poor velocity precision of the AAT data
($\epsilon_{RV} \sim 12$ km s$^{-1}$).
However, many analyses of the dynamics and $M/L$
of this system specifically make use
of the single dispersion ($\sim11$ km s$^{-1}$) derived for Ibata et al.'s database
of 114 stars located at the nucleated 
(perhaps unrepresentative) center.  

Recently, a first analysis of a new AAT data set was presented by, Pe\~narrubia et
al. (2011), based on 1805 Sgr 
members by RV in 6 fields along the major
axis and one side of the minor axis of Sgr reaching to $\pm 4^{\circ}$
along the major axis.  This data analysis and modeling focuses primarily on
whether rotation is found in Sgr.   

An improved kinematical analysis of the Sgr inner core and attempting to
isolate it chemically and kinematically from M54 was conducted by
Bellazzini et al. (2008).  They analyzed 1152 VLT/FLAMES and Keck/DEIMOS
spectra using RVs and CaT metallicities determinations to isolate the
Sgr core population from that of M54.  While this provides an
excellent analysis of the central core of Sgr ($\sigma_{v} = 9.6 \pm
0.4$ km s$^{-1}$), the limited area, all
within 9' of core, provides just one data point compared to the
larger main body of Sgr.

Of course, understanding the Sgr system --- among the nearest known examples of a minor merger --- 
is sufficient cause for systematic exploration of its core, whether or not it can be considered a suitable 
representative of the dSph morphological class in general.  
Some recent evidence makes a compelling case 
that the primary difference between Sgr and other MW satellites may be principally in its life phase
within what may be a universal evolutionary course for dwarf galaxies.  
In general, the star formation histories of dwarf galaxies in the local universe are similar
for most of cosmic time, regardless of their present morphological type, with the latter
mostly a function of variations in the influence of external mechanisms 
(like ram pressure stripping or tidal effects) within the past Gyr or so
and that give rise to a strong morphology-density relationship
(Skillman et al. 2003; Grebel 2004; Weisz et al. 2011).
According to models by Mayer et al.\ (2001), tidal stirring via close passages around 
large host galaxies can transform star forming, gas-rich, dwarf irregular (dIrr) systems with 
small rotating disks into classical, 
moribund, gas-depleted, pressure-supported dSphs.
The mechanism driving such a metamorphosis is the formation and decline
of a central bar, incited by the dynamical interaction of the satellite with the gravitational
potential of the host, particularly at pericenter passages.  Certainly it is clear that at least
some of the present ``classical'' MW dSphs (e.g., Carina, Leo I), though now devoid of gas, 
were forming stars in the not too distant past (the past few Gyrs; Smecker-Hane et al.\ 2009;
Gallart et al.\ 2007) ---  just like Sgr \citep{siegel07}.  Interestingly, Carina and Leo I may be 
the two classical MW dSphs apart from Sgr with the strongest evidence for tidal disruption
\citep*[Carina:][ Leo I: Sohn et al.\ 2007; Mateo et al.\ 2008]{IH95,kuhn96, majew00, majew04, munoz08}. 

Ironically, despite Sgr's proximity, less is known about the distribution of properties of stars across the 
Sgr core than for other satellites of the MW, for a variety of potential reasons --- e.g., because of
(1) the tendency of observers to avoid it
on grounds that it is perceived to be unusual, as mentioned above, 
(2) the fact that Sgr has only been relatively recently discovered (Ibata et al.\ 1994), as well as, 
probably most importantly,
(3) its intrinsic size at its distance from the Sun means that Sgr spans a large angle on the sky.
Even with large field of view, multifiber spectroscopy it is simply difficult to cover the enormous
size of the Sgr core (with a semi-major axis length of almost 30 degrees).
Nevertheless, some basic knowledge about the Sgr core is, of course, in place.

Ibata et al.\ (1994, 1995) made the first maps showing the extended and elongated nature of the Sgr core
and determined that it was comparable in size and luminosity to the largest of the MW
dSph satellites, the Fornax system.  
These authors also showed that, like Fornax, the Sgr system has its own retinue of globular
clusters, including M54, a large globular cluster that is either the very heart, or {\it at} the very 
heart, of the dSph (see discussion by Bellazzini et al.\ 2008; Monaco et al. 2005; but also cf.
Siegel et al. 2011).
Other globular clusters in the Galactic
halo farther from the Sgr core may also be debris from the original satellite (see review by
Law \& Majewski 2010b).

Majewski et al.\ (2003; hereafter PAPER I) assembled the first all-sky
view of the Sgr dwarf galaxy core along
with its tidal tails by identifying constituent M giant stars within the 
final release Two Micron All-Sky Survey (2MASS; Skrutskie et al.\ 2006) point source catalog.
M giants, which are copious in the relatively metal rich ([Fe/H] as high as 
$-$0.4; Layden \& Sarajedini 2000) Sgr populations, are separable from M dwarfs
with $J$,$H$,$K$ photometry (Bessell \& Brett 1988), and in any case the former well outnumber
the latter to 2MASS magnitude limits (e.g., $K_s \sim 13-14$).  The 2MASS M giant 
sample, which is more robust to the differential reddening effects
that have made optical studies of Sgr core difficult, yielded the first ``clean'' view of the
entire main Sgr body.
From the distribution of the large population of
M giants, the morphology of the Sgr core could be more accurately
fit, and Majewski et al.\ showed that it could be described by a King profile but with,
at the largest radii, a break in the King profile due to the presence of its tidal tails.
Apart from the fact that Sgr shows a small density cusp at its very
center, which may or may not (see Monaco et al. 2005) be due to the
density enhancement provided by the superposition of M54,
the overall King+power law break profile of Sgr looks identical to those previously observed 
in other ``classical'' Galactic dSphs (see Fig.\ 13 below); however, the source of these profile
breaks in other dSph systems have 
been debated as due either to prodigious tidal mass loss or as bound populations 
expected for the high inferred $M/L$ in these systems (e.g., Munoz et al.\
2006, 2008; Sohn et al. 2007; Koch et al. 2007; 
McConnachie, Pe\~narrubia, \& Navarro 2007).  
It is commonly suggested (e.g., Kroupa 1997; Kleyna et al.\ 1999) 
that discrimination of these two models should be possible 
from the radial velocity (RV) dispersion trend with radius.  Sgr represents the closest, most accessible dSph in the class of those exhibiting King+break profiles, 
but it is also the one example where the break is from {\it certain} tidal disruption.  
Thus it provides a Rosetta Stone by which the dynamical markers of this process
may be empirically established.

In the present paper we undertake the first large, wide-field {\it and} 
high resolution ($R\sim15,000$) spectroscopic survey to sample across 
the larger area encompassing the core body of the Sgr system.  
With this survey, we explore the velocity 
distribution of 2MASS-selected
K and M giant stars across the
face of this core, from its very center out to the trailing tidal tail and along 
the major, minor and other symmetry axes of the system.  Thus, with this survey
we provide kinematical 
information for Sgr that is comparable 
in extent to datasets existing for other Milky Way dSphs.  This enables us to put 
Sgr in a dynamical context 
with these other systems, and provide
a baseline for comparison to a dSph known to be undergoing severe tidal disruption.  
Our study focuses on 
two primary observables to provide this context:
(1) Within the distribution of radial velocities (\S3)
we look for signatures of rotation and show that Sgr, like other dSphs, has
no significant rotation signature (\S4).  
(2) The velocity dispersion profile of Sgr is shown to be more or less flat, which
is the same trend found in all other dSphs (\S5).
Thus, in these two properties, Sgr is found not to be unusual compared to 
other dSphs, despite
its current state of enduring extreme mass loss (see, e.g., PAPER I; Chou et al. 2007).  
We comment on the potential
significance of this apparent ``normality'' of Sgr as a dSph system for the 
potential state of other dSph systems in \S6.2.

\section{SPECTROSCOPIC SURVEY OF THE SGR CORE}
\subsection{Survey Coverage and Observations}

\begin{figure*} \epsscale{1.0}
\begin{center}
\plotone{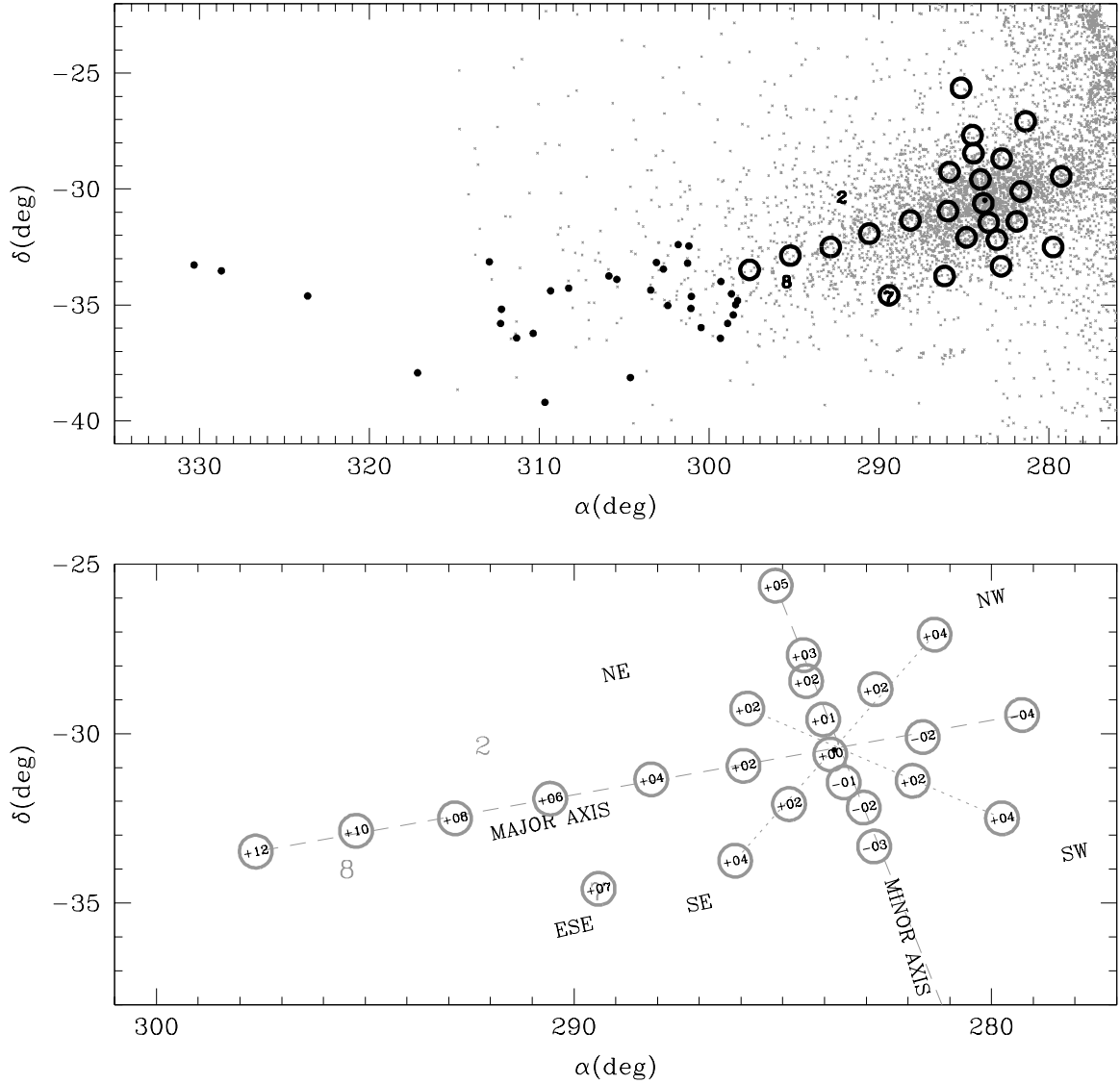}
\end{center}
\caption[]{\label{Sgr_hydra} 
(a) View of the central parts of Sgr and the location of our survey fields.
M giants from PAPER I up to $b = -5\arcdeg$ are shown.
The locations of globular clusters are marked as follows: 
``2'' is Arp 2, ``7'' is Terzan 7, ``8'' is Terzan 8.
Black points are the locations of the additional stars described in \S 5.1
(b) Same as shown in (a) but with fields identified with descriptive labels.}
\label{fig:Sgr_hydra}
\end{figure*}

The present survey systematically samples the core of the Sgr dSph galaxy with target fields selected 
primarily along the major and minor axes.  Additional pointings along diagonal directions
give greater leverage in mapping the kinematical and chemical distributions of the dwarf. 
Fields were selected to explore the major axis at $\sim$2$\arcdeg$ intervals, 
the minor axis at $\sim$1$\arcdeg$ intervals, and the diagonals at +2$\arcdeg$ and +4$\arcdeg$ radial separations (Fig.~\ref{fig:Sgr_hydra}).  
The precise field centers were allowed to shift by a few arcmin to
maximize the number of observable targets, as described below.

\begin{figure} \epsscale{1.11}
\begin{center}
\plotone{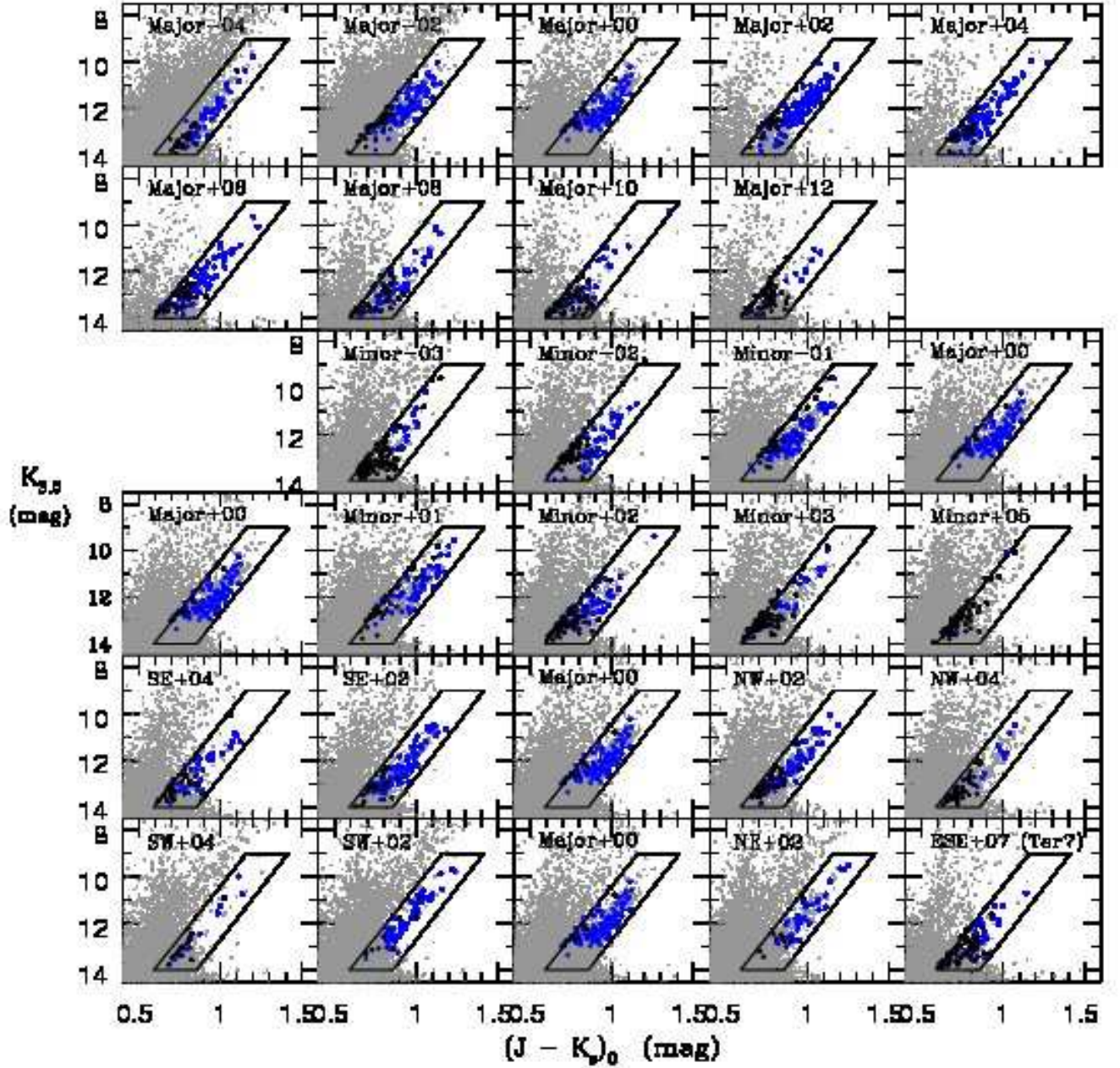} 
\end{center}
\caption{\label{CMDdered}Dereddened 2MASS color-magnitude diagrams for each observed field.
The box delimits the selection criterion used to select Sgr RGB candidates for targeting. 
Sgr member stars based on the RV as described in \S 3 are shown in
black, non-members are shown in blue.     
The panel corresponding to the center field (i.e., ``Major+00'') appears several times in the figure to complete
sequences of panels corresponding to different cross-sections across
the Sgr system. (Degraded image for astro-ph)
}
\end{figure}

The potential target stars were selected to be late type giants as identified in dereddened
$(J-K_s, K_s)$ color-magnitude diagrams from the 2MASS catalog.  
The 2MASS photometry was dereddened using the Schlegel et al.\ (1998) maps, with their
$E(B-V)$ values converted to $E(J-K_s)$ and $A(K_s)$ using the relations given in
PAPER I.
The maximum inferred mean reddening in any field was $\left<E(J-K_s)\right> = 0.224$ in Minor$+$05.
Stars that may belong to the Sgr red giant branch (RGB) were selected using 
the following color-magnitude region:
\begin{itemize}
\item $9 < K_{s,0} < 14$ 
\item $K_{s,0} > -8.8 \, (J-K_s)_{0} + 19.1$ 
\item $K_{s,0} < -8.8 \, (J-K_s)_{0} + 21.6$ 
\end{itemize}
\noindent  This selection was made to isolate the Sgr RGB in the Sgr
core (e.g., field Major$+$00) while minimizing the
contamination from Galactic stars, as shown in Figure \ref{CMDdered}, though
distance and metallicity variations can cause the mean RGB color and
magnitude to shift which was not accounted for in this study.

Spectroscopic data were obtained using the Hydra spectrograph on the Blanco 4-m telescope 
at CTIO on the nights of 19--22 July 2003 and 1--7 August 2003 under
mostly clear skies with good seeing 
(estimated 1.2\arcsec).  Data were obtained using 
132 fibers that are simultaneously dispersed onto a $2048\times4096$ pixel CCD 
(SITe400mm) using the 380 grating (1200 lines mm$^{-1}$) 
and with the fiber ends viewed by the spectrograph through 
a 100$\mu$m slit plate to improve the resolution to
0.68 \AA ~per resolution element ($R \sim 15,000$).  
The spectral range covered was 7740--8740 \AA\  and we typically 
exposed for 60 minutes per field to obtain spectra with signal-to-noise 
ratios ($S/N$) of 5 for stars 
at the faintest limit
of $K_s \sim 14.0$.

Potential Sgr stars were weighted by magnitude and input 
to the {\tt hydra\_assign} program that is used to design fiber configurations
for the Hydra multi-object spectrograph on the Blanco 4-m telescope 
at the Cerro Tololo Inter-American Observatory (CTIO).
The {\tt hydra\_assign} parameters were set to maximize the
number of Sgr candidate stars selected, and this was done by 
testing the total weights for assigned fibers at each field centering for
a spatial search grid of $15\arcmin \times 15\arcmin$ 
with 1$\arcmin$ steps centered on the nominal 
field center. 
The centers of the actually observed fields using this optimization 
plan are found in Table \ref{FieldCent} and shown in Figure \ref{Sgr_hydra}.

Along with our target fields, we also observed a selection of radial
velocity standard stars.  For each of these RV standards, we obtained
multiple observations, where each ``observation'' entails sending the light of the calibrator
down 2-12 different fibers with slightly differing exposure lengths to
be able to measure the trend of velocity precision with data quality.

\section{RADIAL VELOCITIES AND MEMBERSHIP ANALYSIS}
\subsection{Measuring the Radial Velocities}

Radial velocities were determined using IRAF {\tt fxcor}, which cross-correlates
stellar templates with the object stars based on the methodology of
Tonry \& Davis (1979).  For this study, the stellar
templates were high-S/N RV standard stars.  The region used in the
cross-correlation was 8220--8680 \AA.   
A full discussion of the data reduction and calibration to the IAU standard RV system can be found in 
\citet{Thesis1}; the present data were obtained on the same nights
using the same ``red'' RV standards.

Only stars with $S/N$ of 5 or better were included in the RV analysis.  
However, spectral template mismatches also produce poor results, so an
additional culling based on the Tonry-Davis ratio (TDR $>$ 3) was also
employed. The TDR selection is used over a $S/N$ one because the TDR statistic
also accounts for the
effects of mismatches (e.g., due to
temperature/chemistry differences) of a target star's photosphere with that of 
the velocity template; the TDR therefore
provides a better correlation to the measured velocity errors. 
We determined our velocity uncertainties as fully described in
\citet{Thesis1} and adopting the method of Vogt et al.\ (1995), which
calibrates the uncertainty as a function of the measured TDR.  This
calibration was done independently for each observing run, and we find that our data
are on the standard system to within 0.3 km s$^{-1}$. 
For the dynamical analysis presented in this paper, only
stars with uncertainties of less than 5 km s$^{-1}$ are used.
Velocities were converted from the measured (heliocentric)
values to the Galactic Standard of Rest after
correcting for Earth barycentric motion, assuming a solar motion (in a right-handed, Cartesian coordinate
system) of $(U, V, W)_{\odot} = (10.0,5.2,7.2)$ km s$^{-1}$ (Dehnen \&
Binney 1998), and using the IAU value of $\Theta_{LSR} = 220$ km
s$^{-1}$.  
The full data obtained for each star are listed in Table~\ref{Tab:RVstars}, including the Sgr
field name (col.\ 1), 2MASS ID and
dereddened photometry (cols.\ 2-4), Schlegel et al.'s (1998) derived
$E(B-V)$ for the star's position (col.\ 5), 
Galactic 
coordinates (cols.\ 6 and 7), Sgr longitude, measured along the major
axis (col.\ 8; $\Lambda_{\sun}$ from PAPER I),
the measured RV and uncertainty (cols. 9-10), the TDR (col. 11)
and Sgr membership determination (col. 12; see \S 3.2).

Figure \ref{CMDdered} shows the 2MASS CMDs of each field and the stars observed with
reliably ($\langle\epsilon_{V}\rangle < 5$ km s$^{-1}$) 
measured RVs marked. 
Stars determined to be Sgr members (see below) are shown in blue, and
those found to be non-members in black.  
As may be seen, RV members with a wide range of
RGB magnitudes are sampled, but because we prioritized targets by magnitude,
typically in the more densely populated fields
these members are sampling temperatures primarily
associated with M giant spectral types (i.e., $J-K \gtrsim 0.9$).  For fields that sample lower density
regions of the 
Sgr dwarf the RV members are drawn from the hotter, fainter part of the RGB where 
the luminosity function yields more potential candidates, including K spectral types.

\subsection{Selecting Likely Sgr Members}

\begin{figure} \epsscale{1.11}
\begin{center}
\plotone{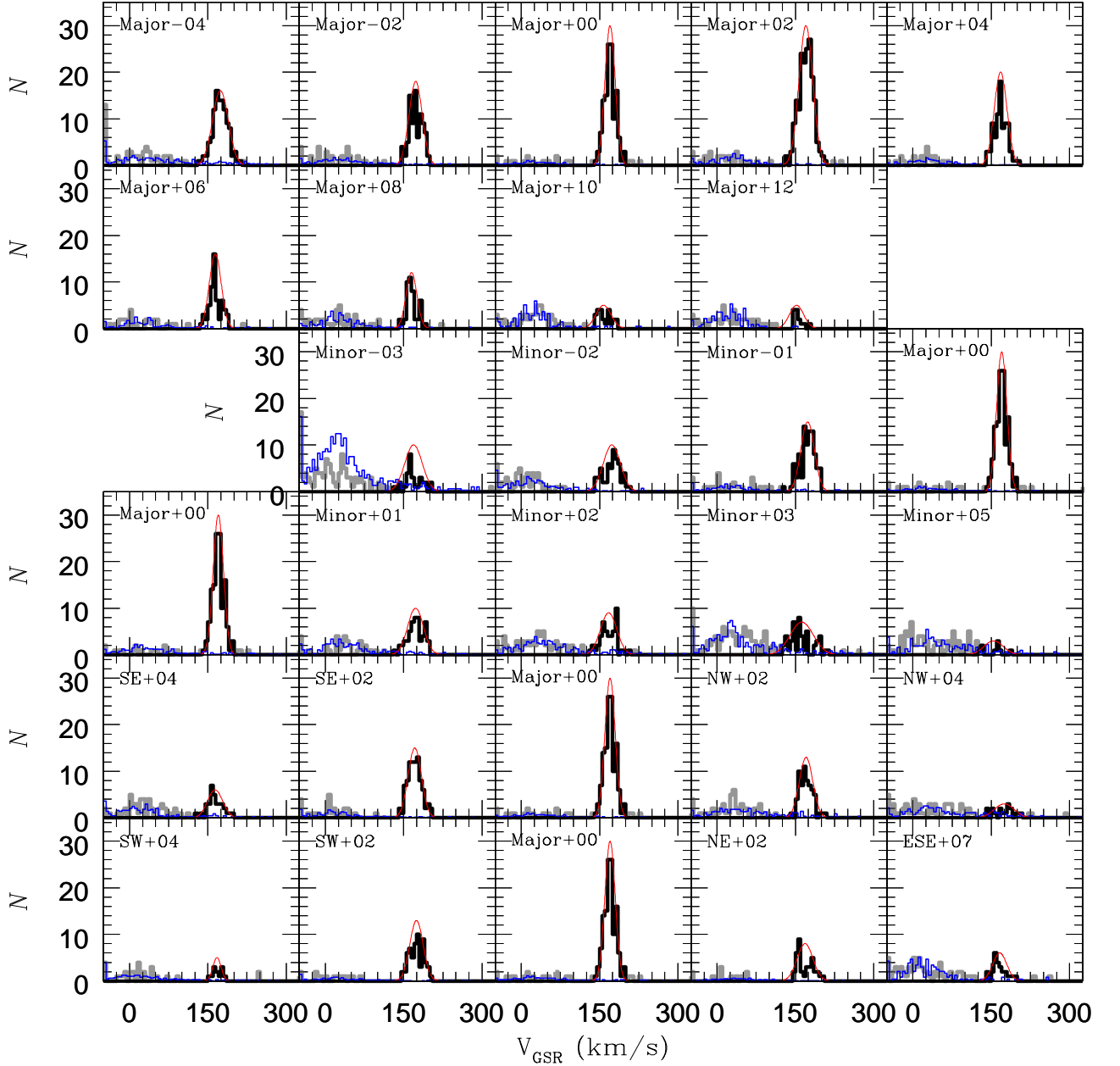}
\end{center}
\caption{\label{RVhists} Radial velocity distributions for each field pointing (grey histograms). 
The lowest probability Sgr members were selected against using a $3\sigma$, iterative
rejection scheme.  
The black histogram shows the stars within $3\sigma$ of the
Sgr mean value.  
The measured means and dispersions of the selected members are
shown as Gaussians (red line), based on data
from Table~\ref{Tab:RVs}.
The blue lines show the results of the Besan\c{c}on Galaxy Model
  (Robin et al.\ 2003)
  for these fields scaled to the number of non-members, and showing
  the generally low level of contamination.
}
\end{figure}

We select stars likely to be associated with the Sgr core based solely on their RVs
with respect to the Sgr mean.
The field-by-field RV distribution of stars
and the ultimate membership selections are shown in Figure
\ref{RVhists}. 
Likely Sgr member stars were selected in each field by using an
iterative 3$\sigma$ rejection technique including measurement 
uncertainties as fully described in \citet{pm93}.
Due to the low numbers of Sgr members in the outer, lower Sgr density fields ($> +02$) 
a pre-selection ($90 < V_{GSR} < 220$ km s$^{-1}$) was 
made to eliminate obvious Galactic contamination, which is centered
around $V_{GSR} \sim 25$ km s$^{-1}$ .  
These membership evaluations for each star are included as the last column
in Table 2.

The net census of RV-classified members and the derived kinematics for
each field 
are compiled in Table~\ref{Tab:RVs}, which shows the mean RV (col.\ 2),
velocity dispersion (col.\ 3), radial distance of the field from the Sgr core
(col.\ 4), elliptical distance of the field scaled to the major axis (col.\ 5), and the
number of ``member'' stars used in the mean RV and dispersion determinations (col.\ 6).
The derived mean, dispersion and total number of members is represented
in Figure \ref{RVhists} as a (red) Gaussian with these properties. 
The actual distributions of Sgr members in each field are Gaussian at
the 99\% confidence level, except for two fields: Major$+$12, which has
few stars, and NE$+$02.

\subsection{Galactic Contamination}

Sgr members selected to this point are those stars falling along the nominal
Sgr RGB in 2MASS CMDs and having radial velocities within a reasonable range of
expected values for the rather rapidly moving Sgr core.  Nevertheless, it is possible
that MW contaminants of 
identical color-magnitude combination might slip
into our ``member'' list if they fall along the extreme tails of the Galactic
field star velocity distribution.  While we cannot identify specific contaminants
without further information, we can estimate the expected number of such stars
using a model for the MW.

To this end, the distribution of velocities in each field was modeled using
the Besan\c{c}on MW model (Robin et al.\ 2003; see Table~\ref{Contamination}).  
For each field, the model output of the Galactic RV distribution for stars that
satisfy our CMD selection (col.\ 2) was scaled to match the number
of RV non-members in our spectroscopic sample for each field
(col.\ 3) according to the ratio of the latter to the total 
CMD-qualifying Besan\c{c}on stars (col.\ 4).  
From these scaled Besan\c{c}on RV distributions (shown as blue lines
in Fig.~\ref{RVhists}) we find the expected 
contamination as those stars falling in the adopted ``RV member'' range (col.\ 5).
Compared to our ``member'' stars in each field (col.\ 6) we find the 
fractional contamination (col.\ 7) to be a
minor effect except in three fields (Minor$-$03, Minor$+$05, 
and
NW$+$04).  This evaluation of a large contamination fraction in these fields
conforms to a simple visual inspection of the scaled Besan\c{c}on RV distributions
in these fields (Fig.~\ref{RVhists}).
We exclude these three fields from further analysis in this paper.

\section{KINEMATICS OF THE SAGITTARIUS DWARF GALAXY CORE}

\subsection{Central and Global Velocity Trends}

The velocity data derived in the previous section can be used to profile the trends of mean
velocity and velocity dispersion across the Sgr core.
First, we compare our 
{\it central} velocity dispersion to that obtained by
Bellazzini et al.\ (2008), who provide high precision measurements of the inner 9' of Sgr and
the globular cluster M54.
We find that our
center field velocity dispersion (Major+00; $9.85 \pm 0.73$ km s$^{-1}$) is consistent with the dispersion
they derive for Sgr {\it after} removing M54 ($9.60 \pm 0.40$ km s$^{-1}$), which is expected given that
our selection of stars is over a relatively
large area. 
As a check on the effects that M54 would have, we also calculated the 
dispersion for the Major+00 field after removing the 33 stars 
in the area covered by Bellazzini et al. (2008; e.g., 9$\arcmin$
radius around M54).  The new dispersion ($9.90 \pm 0.81$ km
s$^{-1}$) shows that M54 has a negligible effect on our derived central velocity dispersion.

\begin{figure*} \epsscale{1.05}
\begin{center}
\plotone{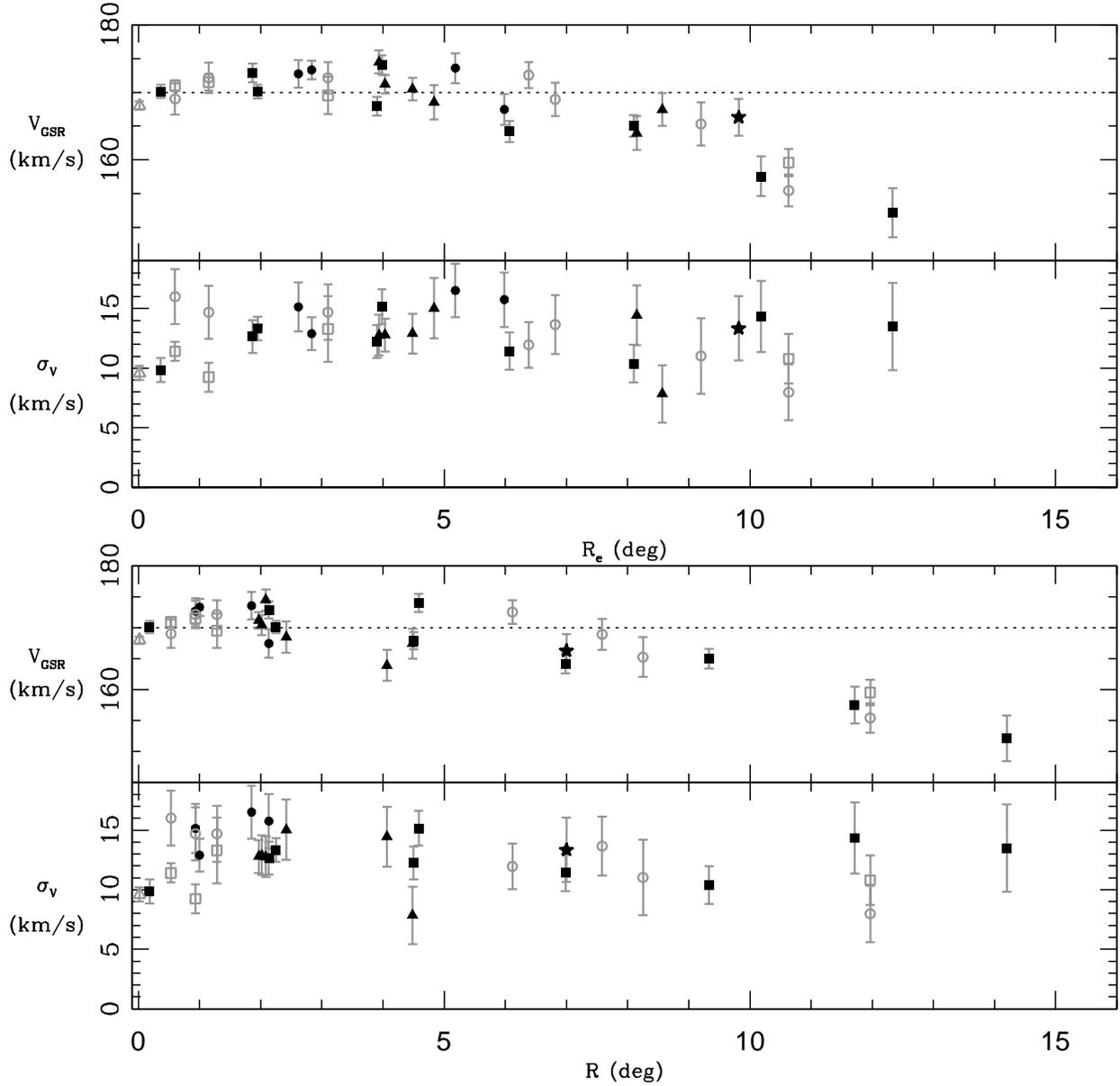}
\end{center}
\caption[]{\label{fig:vGSR_R_Re} Velocity and velocity dispersion
  trends as a function of projected circular ($R$) and elliptical ($R_e$) radii.
Black squares denote fields on the major axis, black circles the minor
axis,  black triangles are from the diagonal fields (NW, SW, NE, SE), and
the black star is the ESE+07 or Ter7 field. 
The fields from Ibata et al.\ are included as grey open circles for the  AAT fields and
open squares for the CTIO fields. The Bellazzini et al.\ field
(Sgr,N) is shown as an open triangle. 
}
\end{figure*}

The only other large-area RV study of the Sgr core, though with lower
velocity precision, is that of Ibata et al.\ (1997).
Most of the Ibata et al. data were obtained using AAT/AUTOFIB,
which resulted in per star uncertainties of $\sim 12$ km s$^{-1}$ (i.e., of the same
order as the Sgr core velocity dispersions).  
This lower precision probably results in the 
inflated near-center velocity
dispersions obtained from these data (visible, e.g., in Figs. 4 and 6, gray
open squares and open circles).  
Additional data were
taken by Ibata et al. with CTIO/ARGUS at a 
higher precision, though the actual resulting
uncertainties are not given by these authors (we estimate them to be $\sim4$
km s$^{-1}$ based on comparison to their repeat observations presented
in their Table 2(b)). 
We find a smaller central velocity
dispersion ($9.85 \pm 0.73$ km s$^{-1}$) than Ibata et al.\ ($11.4 \pm
0.7$ km s$^{-1}$), which, again, is likely explained by the high precision data
of our study, and perhaps underestimated velocity errors for the Ibata et al. ARGUS data.
The Ibata et al.\ data are for stars around the center of Sgr with a couple
outlying fields that fall near the major axis. 
For comparison, their 
dispersions are presented in
our Table~\ref{Tab:RVs} and projected into the trends of our survey in
Figures~\ref{fig:vGSR_R_Re} and \ref{Sgr_hydraA1}, which show a good
correspondence.

\begin{figure} \epsscale{1.05}
\begin{center}
\plotone{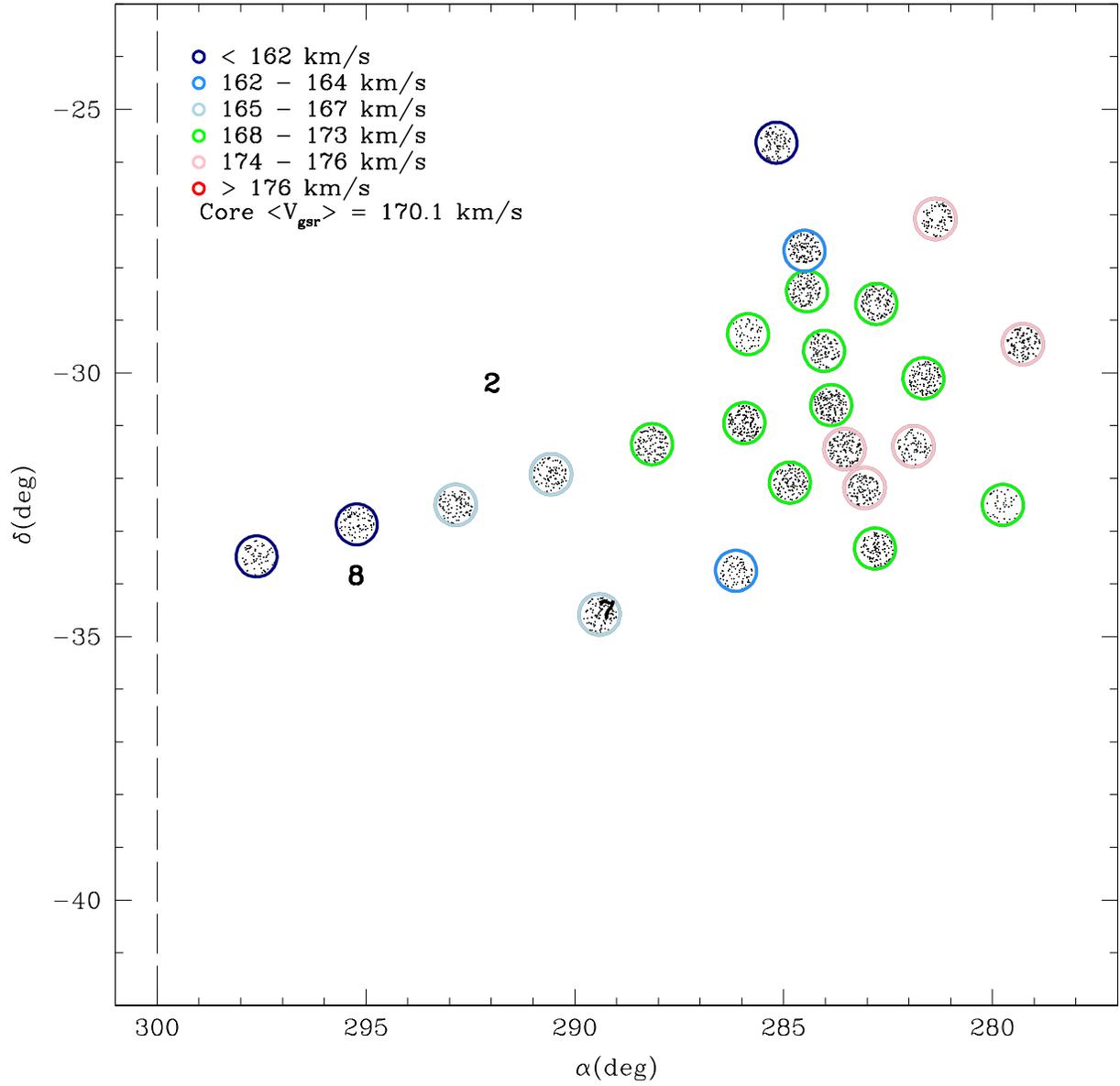}
\end{center}
\caption[]{\label{Sgr_hydra3} View of the central parts of the
Sgr core showing the general trends of mean $V_{GSR}$.  
}
\end{figure}

Figure~\ref{fig:vGSR_R_Re} shows the measured velocity and velocity dispersion 
trends versus both the straight
projected radius
as well as the elliptical radius (i.e., the radius normalized to the semi-major axis along the
elliptical density contours using the shape determined in \citealt{majew03}). 
We find a declining $V_{GSR}$ trend as a function of radius in both
$R$ and $R_e$, however this format is not very sensible for interpreting the  $V_{GSR}$ trend, since it azimuthally
averages over a general velocity trend along the satellite, one seen predominantly along 
the major axis (see below).
The presence of a global $V_{GSR}$ trend is evident by the distribution of mean velocities
color-coded in the survey map shown in Figure~\ref{Sgr_hydra3} and discussed further below.

\subsection{Testing for Rotation}

In addition to looking at global radial trends, by exploiting our areal coverage we can explore trends along four 
different primary axes (major, minor, SW--NE, and NW--SE).  This is particularly helpful for looking for
a signature of Sgr rotation.
Ibata et al.\ (1997) first looked for evidence that Sgr may show rotation along 
the minor axis as an effect of tidal stripping, after claims for
rotation in Ursa Minor dSph were made (Hargeaves et al. 1994). 
Their data showed no evidence of minor axis rotation.
More recently, Pe{\~n}arrubia et al. (2010) predicted a Sgr core rotation as large as 20 km s$^{-1}$ as
a way to ``naturally'' produce a bifurcated leading arm among the Sgr tidal debris.
Such a bifurcation is one explanation for the 
apparent double arms of Sgr debris
in the northern Galactic hemisphere ``field of streams'' (Belakurov et al. 2006).
Our data provide an excellent opportunity to re-verify the Ibata et al. claim for no measured rotation 
along an even greater span of the main body than they explored,
and to test Pe{\~n}arrubia et al.'s prediction 
of a rotating Sgr core.

\begin{figure*} \epsscale{1.05}
\begin{center}
\plotone{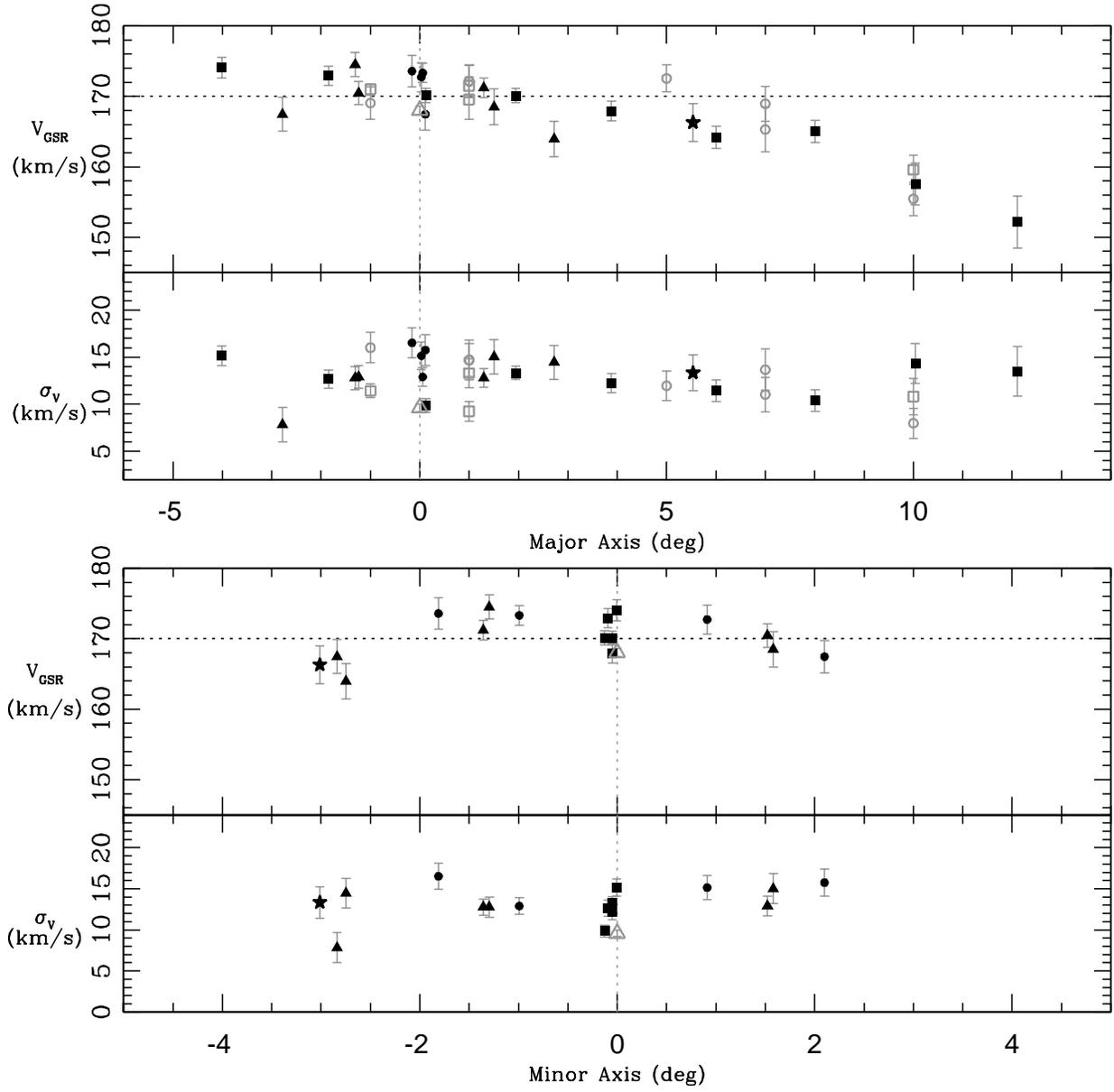}
\end{center}
\caption[]{\label{Sgr_hydraA1} 
Velocity and velocity dispersion as a function of major axis and 
minor axis position. Symbols are same as Figure~\ref{fig:vGSR_R_Re}. 
}
\end{figure*}

\begin{figure*} \epsscale{1.05}
\begin{center}
\plotone{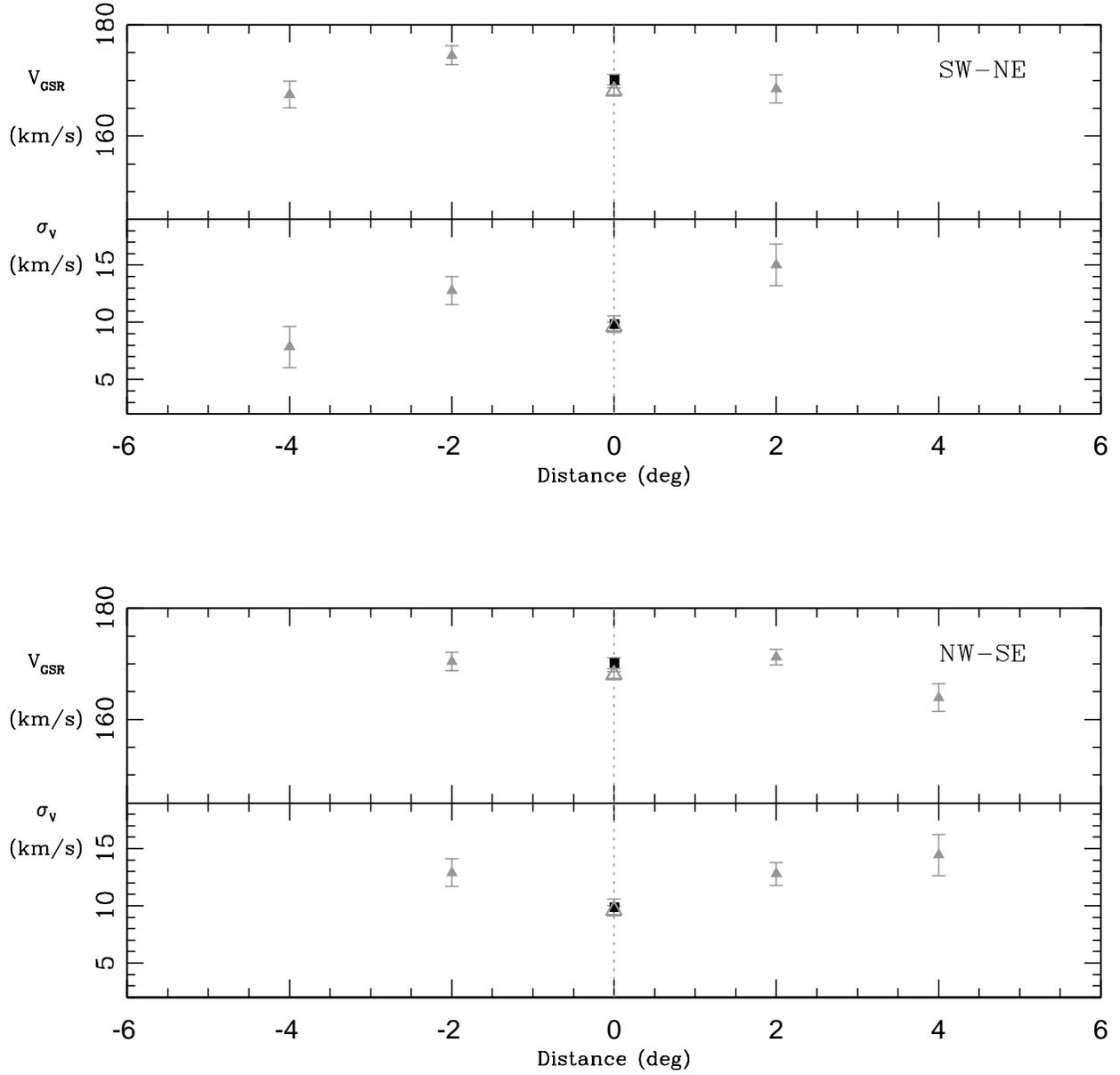}
\end{center}
\caption[]{\label{Sgr_hydraB} 
Velocity and velocity dispersion trends along the diagonal axes 
from SW$+$04 across the core to NE$+$02 (upper two panels) and from NW$+$04 across the core to SE$+$04
(lower two panels).  Symbols and colors are same as Figure~\ref{fig:vGSR_R_Re}.
}
\end{figure*}

To clarify the trends observed, it is beneficial to explore them as a function of position
along the major and minor axes (Figure \ref{Sgr_hydraA1}) 
as well as along the diagonal (Figure \ref{Sgr_hydraB}) axes across the center of Sgr.  
It is clear that our data show no significant gradient at all along
the minor axis, in agreement with the data and suggestion by Ibata et
al. (1997) for a lack of minor axis rotation.
On the other hand, Figure~\ref{Sgr_hydraA1} shows a clear global velocity gradient along
the major axis, which could be a signature of rotation; however the
observed trend is contrary to that proposed by Pe{\~n}arrubia et al.
Their models suggest that there 
should be an {\em increase} to $V_{GSR} \sim 180$ km
s$^{-1}$ along the major axis in the trailing arm direction,
 but we find a {\em decrease} to $V_{GSR} \sim 150$ km
s$^{-1}$.  In addition, given
the large area on the sky spanned by the major axis coverage other possibilities
exist 
to explain these trends besides rotation; these are fully explored in
\S \ref{sec:rot} below.
We see no significant trends in $V_{GSR}$
along the other axes available to our survey (SW--NE and NW--SE), as explored further below.

Pe\~narrubia et al. (2011) have conducted their own radial velocity survey of Sgr specifically 
to look for the Sgr rotation predicted by their earlier model and thereby test whether rotation can 
explain bifurcation seen in the Sgr tails.  Similar to our findings here, these authors
only find an RV trend along the major axis, and no evidence of minor axis rotation, though
their coverage in more limited and only covers one side of the Sgr minor axis. 
Because the individual radial velocities from Pe\~narrubia et al. (2011) have not been published, a
detailed analysis and comparison between our data sets is not yet possible.


\subsection{Modeling Large Area Orbital Effects\label{sec:rot}}

To investigate in detail the kinematical trends seen within the Sgr core data, it is necessary 
to distinguish 
the effects of potential intrinsic rotation from global trends due to the
motion of Sgr along its orbit, which, due to Sgr being observed over a large expanse of sky, leads to 
strong radial velocity variations as observed from Earth.

\begin{figure} \epsscale{1.05}
\begin{center}
\plotone{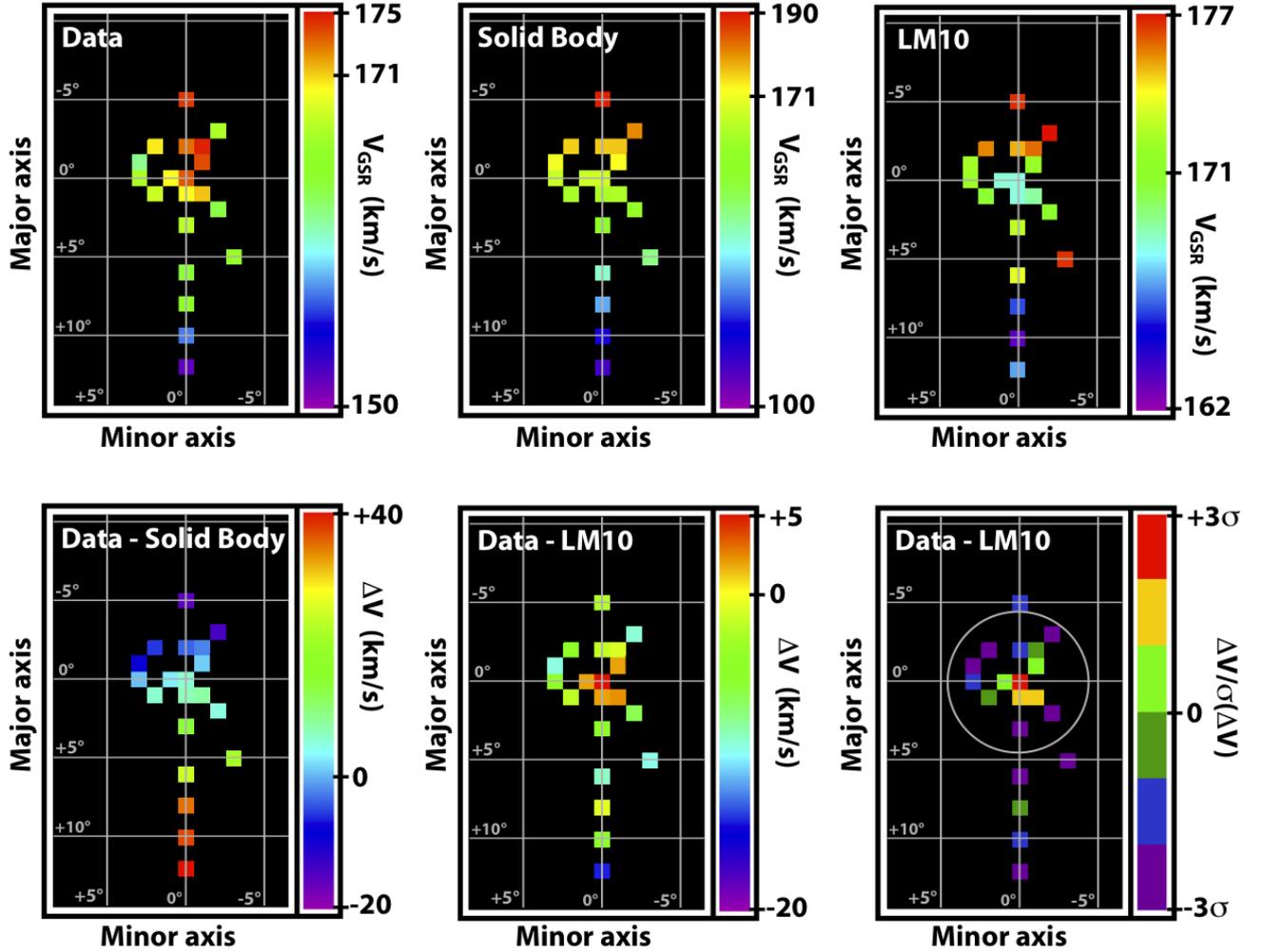}
\end{center}
\caption[]{\label{Sgr_2Dmap} Top row: Velocity maps for the observational data, a solid-body model moving with the space velocity of
Sgr determined by LM10, and the $N$-body model of LM10.  Sampling of the models is chosen to match the observational
data.  Bottom row: Residual velocity difference between the observational data and the solid-body and $N$-body
models.  The bottom right panel shows the significance of positive/negative residuals between the observational data
and the $N$-body model.  The grey circle in the bottom right panel indicates the $\sim 4.5^{\circ}$ radius at which unbound stars are expected to constitute
$\sim 50$\% of the stars in a given field based on the LM10 model. }
\end{figure}

We therefore 
construct a two-dimensional velocity map by binning the velocity
data into 1 deg$^{2}$ bins in the Sgr 
core coordinates,
and compare this map to theoretical models.
As shown in Figure~\ref{Sgr_2Dmap} (upper left panel), the observational data
show a 
statistically significant
velocity gradient along the major axis of Sgr, with
$v_{\rm GSR} \approx 174$ km s$^{-1}$ in the Major-04 field and 
$v_{\rm GSR} \approx 152$ km s$^{-1}$ in the Major+12 field.  
Such a trend can be expected even for a non-rotating model however, and
reflects the changing projection of the space velocity of Sgr stars
onto the line of 
sight with angular position across the dwarf.  In Figure~\ref{Sgr_2Dmap} (upper
middle panel) we compare 
to the velocity trend expected if we were to adopt the simplest possible model
for Sgr; namely an extended solid-body translating with space velocity
$(U,V,W)=(230, -35, 195)$ km s$^{-1}$ (see Law \& Majewski (2010b); LM10 hereafter),
for which all variation of the radial velocity across the body is due 
to projection effects.
The trend of radial
velocities along the major axis in this simple
solid-body model closely resembles the trend in the
observational data,
although the magnitude of the end-to-end velocity differential is
significantly larger ($\sim 90$ km s$^{-1}$) than observed
(Figure~\ref{Sgr_2Dmap}, lower left panel).

A solid-body model is a poor descriptor of real, extended Galactic 
satellites however,
because such satellites respond to the Galactic gravitational 
potential and unbound stars have streaming
motions along the tidal streams.
A more physically motivated comparison may be made to the
$N$-body model of Sgr presented by LM10; this model assumes that Sgr  
was originally an isotropic (i.e., non-rotating) Plummer sphere
and follows the orbits of the $10^5$ individual tracer particles over 
the interaction history of the dwarf, thereby accounting for both
projection effects and varying space velocities throughout the bound 
Sgr core and unbound tidal streams.\footnote{While the LM10 model was not constrained to match the Sgr
  core in detail, its accuracy suffices for the present analysis.
Due to numerical limitations of the $N$-body technique though, the 
LM10 model actually overshoots the current location of Sgr by $\sim 2^{\circ}$
along its orbit.  We manually adjust the LM10 model back along its orbit to correct for this.}
As illustrated in Figure~\ref{Sgr_2Dmap} (upper right panel), the LM10 model also
shows a significant velocity gradient along the major axis, but with an amplitude more
similar to that of the observational
data.  Subtracting the LM10 model velocity field from the observational 
data, we construct a map of the radial velocity difference between the observational data
and the non-rotating LM10 theoretical model (Figure~\ref{Sgr_2Dmap}, lower middle
panel).  In this velocity difference map, the remaining major-axis velocity gradient is no longer
apparent, suggesting that the gradient in the 
observational data is due 
to orbital streaming motions rather than intrinsic rotation in the dwarf.
Combining the observational uncertainties in the mean radial velocity 
for each field in quadrature with 
the corresponding uncertainty from the LM10 model fields, 
it is possible
to convert the velocity difference map to a map of the statistical 
significance of deviations from the LM10 model (Figure~\ref{Sgr_2Dmap}, lower right panel).  
No obvious trends of positive/negative
residuals are apparent, although we note that, intriguingly, the majority of 
observational fields with $r \gtrsim 3^{\circ}$ from the Sgr core have velocities lower than the LM10 model by
$\sim 2-3 \sigma$ (i.e., $\sim 5-10$ km s$^{-1}$).  
The interpretation of this feature is uncertain, but the $r \sim 3^{\circ}$ 
boundary may indicate the transition from predominantly-bound to predominantly-unbound
stars (expected to occur at $r \sim 4.5^{\circ}$ in the LM10 model),
for which streaming motions might be
slightly different than predicted.

\begin{figure} \epsscale{1.05}
\begin{center}
\plotone{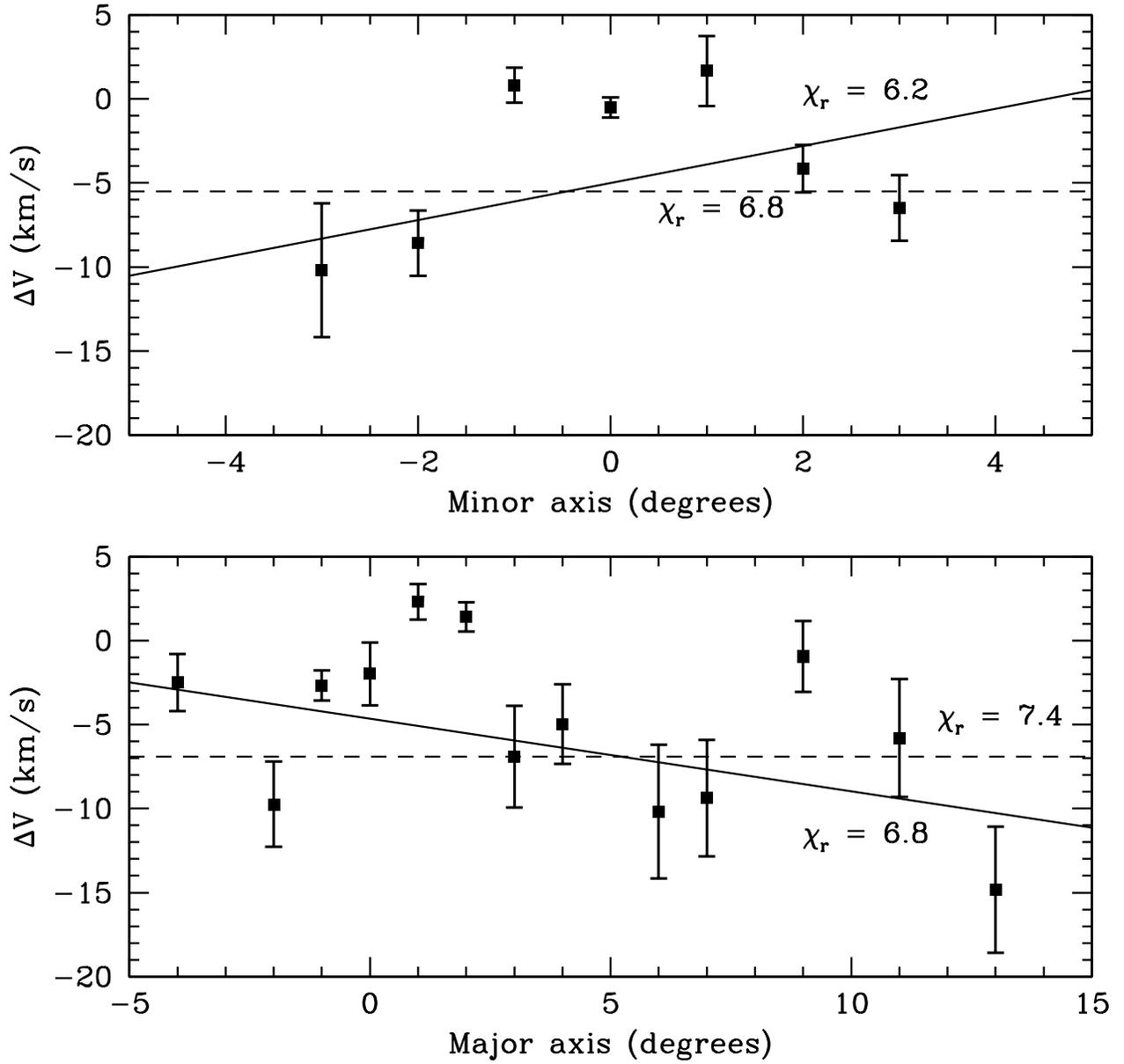}
\end{center}
\caption[]{\label{Sgr_trends} Linear least-squares fits to the velocity residuals (observational data minus LM10 $N$-body model)
along the major (bottom panel) and minor (top panel) axes.  Values for the square root of reduced $\chi^2$ for
each fit are shown.
 }
\end{figure}

In Figure~\ref{Sgr_trends}, we evaluate the significance of the velocity residuals 
remaining along the major and minor axes after subtracting the LM10 model from the observational data.
The residuals are best fit with a velocity gradient declining slightly 
along the major axis (0.5 km s$^{-1}$ deg$^{-1}$) and minor axis (1.2 km s$^{-1}$ deg$^{-1}$),
but the reduced $\chi^2$ of these fits are not significantly better than obtained with a constant velocity model.

\begin{figure} \epsscale{1.05}
\begin{center}
\plotone{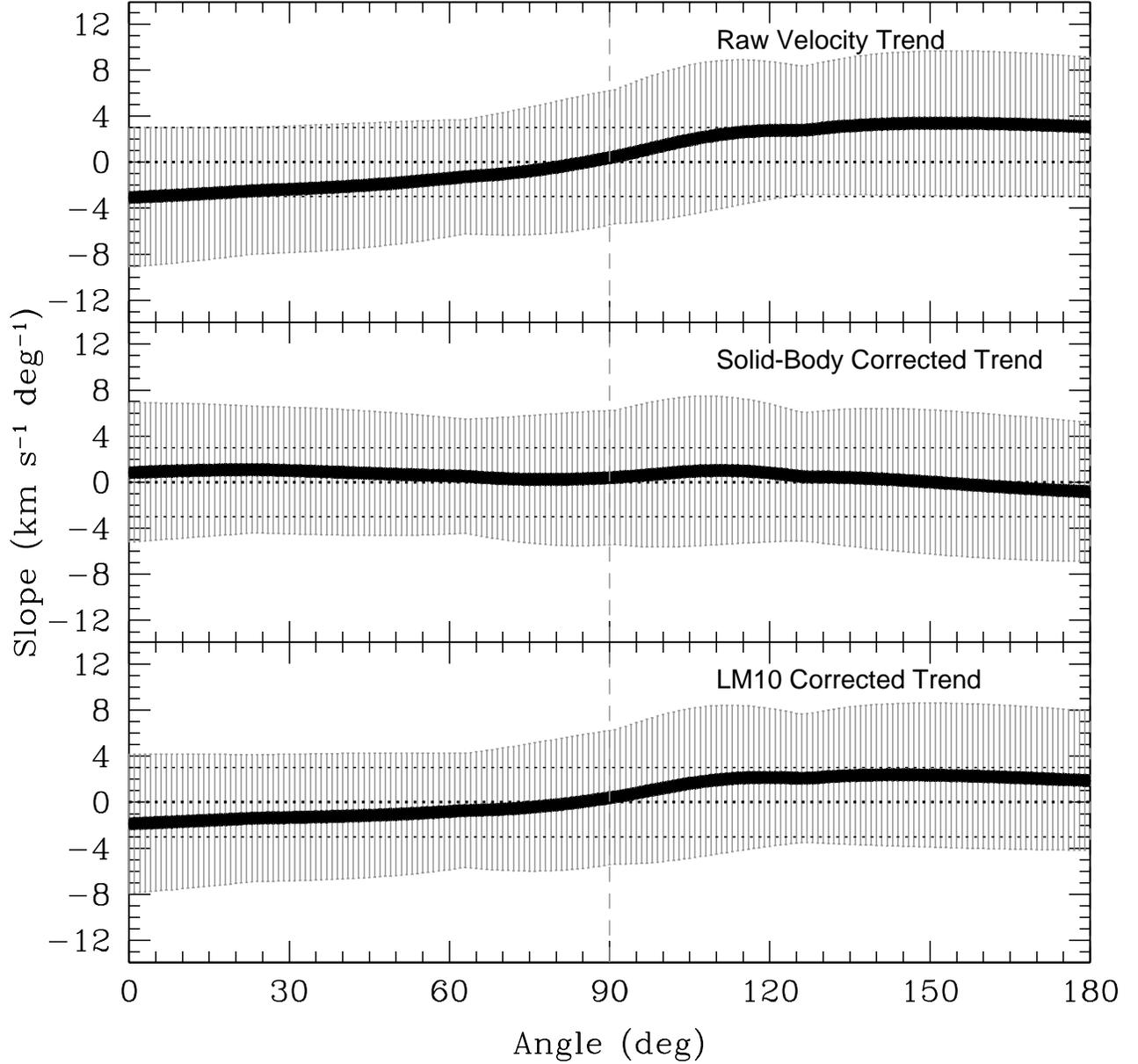}
\end{center}
\caption[]{\label{fig:angle} To test for various effects of galaxy rotation,
  we projected all stars along the major
  axis (Angle - 0) and determined the slope (black points) and
  dispersion (grey bars).  We then repeated these projection on axes
  in $1^{\circ}$ increments through $180^{\circ}$ (back to the Major
  Axis). Minor axis is denoted by the dashed line at 90$^{\circ}$
  (a) Observed trend with no corrections.  (b) Observed trend
  corrected by including the effects of solid-body orbital motion.
  (c) Observed trend corrected by including the modeled dynamical
  effects from Law \& Majewski (2010) model for the global trends of Sgr.
 }
\end{figure}

As another 
way to look for an axis of rotation, we projected the mean velocities of
all fields onto variously oriented axes through the Sgr center.
We projected the stars onto axes rotated every degree starting with the major axis, 
and fit the velocities with a linear slope (0$^{\circ}$ and 180$^{\circ}$ are along the
major axis and 90$^{\circ}$ is along the minor axis).  
We plot these fitted slopes as a function of axis angle in Figure~\ref{fig:angle}.
This analysis (Figure~\ref{fig:angle}) shows no evidence of minor axis rotation
(significant slope across the galaxy)
with either model, again confirming the results of Ibata et al.\ (1997).  
After subtraction of either the solid body or the LM10 models, 
there is no significant rotation seen along {\it any} axis.  At most possibly a slight trend 
mostly on the major axis is seen.

While it is not possible to completely rule out rotation at the level 
of a few km s$^{-1}$ (given observational uncertainties and
uncertainties inherent  
to the simple the solid-body model and fact that the LM10 does not try to
explicitly fit the Sgr core), 
there is no compelling evidence for statistically significant rotation 
about either the major or minor axis (or any other axis)
in the central $\sim 6^{\circ}$ of the Sgr dwarf.
In particular, {\em we find no evidence} that Sgr has a strongly rotating 
($\sim 20$ km s$^{-1}$ amplitude) core, as predicted by
Pe{\~n}arrubia et al.\ (2010) for disk models of the Sgr progenitor 
that are capable of producing a bifurcation in the Sgr leading tidal stream.
This point was previously made based on a simpler treatment of these 
same data by {\L}okas et al. (2010a) and using different data in Pe{\~n}arrubia et al.\ (2011).

\section{VELOCITY DISPERSION PROFILE IN THE PRESENCE OF DARK MATTER
AND TIDAL DISRUPTION}

\subsection{Supplementary Data Bridging the Sgr Core and Tails}

Because velocity dispersion profiles and how they bear on the
mass profile of a dSph have become central to understanding the
total mass and dark matter content of these systems, and because 
we wish to provide as similar a portrait of the dynamics of
the Sgr core as we have for other dSphs, we have sought to append
the dispersion profile of the Sgr system with data even farther from
the Sgr core than provided in Figure~\ref{fig:vGSR_R_Re}.
Thus, we have obtained additional spectra for a small sample of
stars (shown in Figure 1) that bridge from the data shown in Figure~\ref{fig:vGSR_R_Re} 
(which reach $0.47 r_{lim}$, where $r_{lim}$ is the limiting radius
of a King profile fit to the density profile of the Sgr core, found by
Majewski et al. 2003 to have the value $r_{lim} = 1801$ arcmin)
to those velocity data previously published for unbound stars in the trailing 
arm by Majewski et al. (2004).

These supplementary data, which focused on the calcium infrared triplet
spectral region, were obtained with the Modular Spectrograph 
and a 1200 line mm$^{-1}$ grating on the DuPont 2.5-m telescope
on the nights of UT 2003 Sep 19-20.
The stars were selected as 2MASS M giants of the expected distance of
Sgr (see Majewski et al. 2003) and lying near the semi-major axis of Sgr on its 
trailing side and cover the outer parts of the Sgr core, from 12.7$^{\circ}$
(i.e., $0.44r_{lim}$) to 48.2$^{\circ}$ (i.e., $1.6r_{lim}$). 
The positions of the selected stars are shown in Figure 1.
Radial velocities were evaluated using the methodology discussed in
Majewski et al.\ (2004; PAPER II), 
but --- because these data were obtained with a slit spectrograph rather than
with fibers --- with the addition of a post-processing correction to each velocity that accounts 
for errors in the centering of each star on the slit (see discussion of this procedure in Sohn et al. 2007).
These errors --- typically amounting to a few km s$^{-1}$ --- were evaluated by
measuring, via cross-correlation against a template of twilight sky, the relative
position of telluric absorption features, including the A band and the series of
H$_2$O and OH molecular lines near 8200 \AA\ , which directly map
the stellar slit-centering error.
Based on the measured RVs in multiply observed stars, we determine that these
spectra, spanning 7700-8700 \AA\ at a resolution of 1.7 \AA\ per resolution
element, yield RV  uncertainties after the telluric line corrections
of effectively 2 km s$^{-1}$ --- i.e., well-suited to
measuring $\sigma_v$'s and matching well the precision of the
data we collected using Hydra (presented in Table 2).

The data for these supplementary stars are provided in Table
\ref{modspec}.
The full RV data for each star including 2MASS ID and
dereddened photometry (cols.\ 1-3), Schlegel et al. (1998)-derived $E(B-V)$ (col.\ 4),
Galactic 
coordinates (cols. 5 and 6), Sgr $\Lambda_{\sun}$ coordinate (col.\ 7; see \S 2.1),
RV and uncertainty (cols.\ 8-10), 
height of the cross-correlation peak (col.\ 11), data quality
(col.\ 12; see Paper II 
for definition of this quantity, which goes from 0 [worst] to 7
[best]), and membership (col. 13; have RVs consistent with being members of the most recently detached part of the trailing arm). 
A simple scan down the $V_{GSR}$ column of Table~\ref{modspec} shows that
only about 5 of the 37 stars observed with the DuPont have RVs inconsistent
with being members of the most recently detached part of the trailing arm
(though it is possible that a few 
of the stars could be Sgr stars from other wraps of tidal 
debris, particularly at the largest separation from the Sgr center 
--- see, e.g., PAPER II). 
 Grouping the remaining 32
stars into 3 additional radial bins (e.g., Figure~\ref{Sgr_hydra_kunkel}) enables us to follow the 
RV dispersion profile of Sgr to the King limiting radius (30$^{\circ}$)
of the King profile fit to Sgr in PAPER I.

\subsection{Velocity Dispersion Trends}

The distribution of the velocity {\it dispersion} as a function of radius 
(either strictly circular or elliptical) shown in
Figure~\ref{fig:vGSR_R_Re}
 is the typical way that dSph dynamics and mass profiles are explored.
As may be seen, the velocity dispersion of Sgr remains more or less flat 
to the limit of our study, albeit with a colder dispersion in the very center.  
This cold point is not seen in the poorer resolution AAT data of Ibata et al. (2007),
but is seen in their better precision ARGUS data and confirmed also by the
high resolution Bellazzini et al. (2008) data (see \S4.1).  The upper panel
of Figure 6, which shows the data only for fields along the major axis of Sgr, 
demonstrates the cold point most clearly.

A single field, Minor$+$03, shows an anomalously high velocity dispersion, 
by almost a factor of two
compared to other fields nearby or symmetrically placed around the Sgr
main body.  Though the contamination of this measurement by MW
stars is not estimated to be abnormally high (Table~\ref{Contamination}), inspection of 
Figure~\ref{RVhists}
 suggests that the actual situation may be worse than this, in that, 
like the fields already thrown out of our analysis (Minor$-$03, 
Minor$+$05, 
and NW+04), the ``Sgr RV peak'' in Minor$+$03 is the
most overlapped by the distribution of
MW field stars in the same field.  It would not be surprising 
to be concerned about Minor$+$03 if Minor$-$03 is untrustworthy, given the 
likely similar Sgr density in these symmetrically placed fields, 
but an even higher expected MW presence
in Minor$+$03, which is located even closer to the Galactic mid-plane.
Thus, we put no confidence in the measured velocity dispersion in
the Minor$+$03, and exclude it from further analyses.

\begin{figure*} \epsscale{1.05}
\begin{center}
\plotone{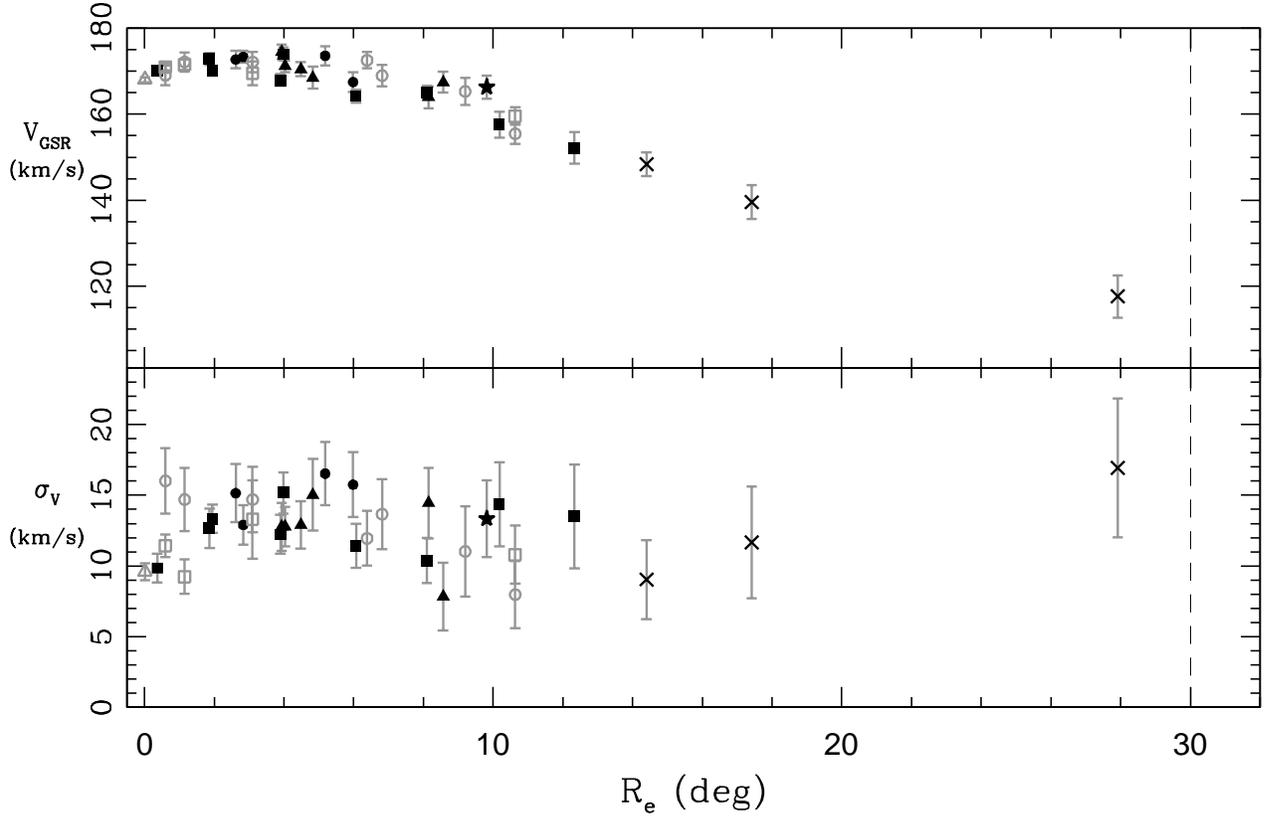}
\end{center}
\caption[]{\label{Sgr_hydra_kunkel} 
Velocity and velocity dispersion
trends as a function of projected elliptical ($R_e$) radii, including
supplemental data from \S 5.1.  The new measurements are shown as
black crosses, and the King limiting radius derived in PAPER I is
shown as a dashed line at $\sim 30^{\circ}$.
As in Figure~\ref{fig:vGSR_R_Re}, black squares denote fields on the major axis, black circles the minor
axis,  black triangles are from the diagonal fields (NW, SW, NE, SE), and
the black star is the ESE+07 or Ter7 field. 
The fields from Ibata et al.\ are included as grey open circles for the  AAT fields and
open squares for the CTIO fields. The Bellazzini et al.\ field
(Sgr,N) is shown as an open triangle. 
}
\end{figure*}

We conclude that there is no evidence for ``sub-structure'' or features in
the dispersion profile --- other than the central cold point.
The velocity dispersion
remains more or less constant as far as it can be probed,
even when we include
the DuPont data at farther radii (\S5.1), as shown in
Figure~\ref{Sgr_hydra_kunkel}.
Obviously Figure~\ref{Sgr_hydra_kunkel} maps the Sgr dynamics from the
bound inner regions of
Sgr out into the regions dominated by unbound tidal debris, wherever that
occurs.
And yet no obvious evidence for any change in the profile due to that
transition can
be discerned.

Similar velocity dispersion profiles to that mapped here for Sgr
have been seen in the modeling of the cores of dark matter-dominated dSph
galaxies
experiencing tidal disruption by Mu\~noz et al.\ (2008).  Those models were
specifically designed to match the properties of the Carina dSph, which in
density distribution and velocity dispersion profile greatly mimics what is
observed in Sgr.  As shown in those models, the flat velocity dispersion
is a reflection of the presence of unbound stars, and, as shown there, the
transition from a system dominated by bound particles to one dominated
by unbound particles generally shows no feature in the velocity dispersion
profile when all stars in the simulations are used, and a
sigma-clipping rejection, similar to
the one used for real data, is applied.\footnote{If one studies the
behavior of the bound
and unbound components separately, then the difference in behavior is
evident, with the
bound part decreasing with radius and the unbound part remaining flat or rising depending
on the specific model, a phenomenon also shown by
Klimentowski et al. (2007).}

This trend of flat line-of-sight velocity dispersion profiles to the limit of
where kinematical data exist is also observed for all other
``classical'' MW dSphs
(Mu\~noz et al. 2005, 2006; Walker et al. 2006, 2007; Koch et al. 2007;
Mateo et al. 2008;
see Figure~\ref{fig:all_disp}).
Given the predominance of unbound debris well inside the King limiting
radius in the case of Sgr (e.g., see \S4.3.3 of Majewski et al. 2003), the overall similarity
of its velocity dispersion profile to those of other dSphs over similar radii
is either an important structural clue or an unfortunate, confounding coincidence.
The flatness of the dispersion profiles in other dSphs has been interpreted
as evidence that mass in these systems does not follow the light
(e.g., Kleyna et al. 2002, {\L}okas et al. 2005, Mashchenko et al. 2005,  
Gilmore et al. 2007; Koch et al. 2007; Walker et al. 2007)
and moreover, that they are surrounded by larger dark matter
halos. However, that a system like Sgr, so clearly undergoing tidal stripping,
shows an identical behavior in its velocity dispersion profile,
even including that portion of the system almost surely dominated by its
unbound stellar component, provides us, at minimum, with a cautionary example
regarding the derivation and interpretation of dSph mass profiles and assumptions
about the extent of the bound component.

\section{DISCUSSION}

\subsection{Summary of Sgr Core Properties Found Here}

We have conducted the first large scale systematic
survey of the
Sgr dSph main body with broad azimuthal coverage.  
Using over 1200 member stars, we are able to
investigate evidence for effects that would be seen over
large areas, such as rotation and global velocity
dispersion trends.
A complementary study of the metallicity distribution across Sgr using these same spectra
will be included in a
future contribution.

The primary results from this paper may be summarized as follows:

1. The Sgr central velocity dispersion is found to be $9.9 \pm 0.7$ km s$^{-1}$,
in agreement with the dispersion
($9.6 \pm 0.4$ km s$^{-1}$)
 found by Bellazzini et al. (2008)
after removing
stars associated with M54 (e.g., Figure~\ref{fig:vGSR_R_Re}, \S 4).  This central velocity dispersion
is not atypical among the ``classical'' dSph galaxies (see Table~\ref{Tab:dwarfs}).
 
2. Within the distribution of radial velocities, we find no signatures 
of rotation along the minor axis of Sgr.  With a detailed investigation of
trends along the major axis, we find evidence for at most a small trend
($\le 4$ km s$^{-1}$ deg$^{-1}$), though even this small effect can be 
explained without needing intrinsic rotation, as shown by various adopted models of 
an extended Sgr system moving along its orbit.
Thus, we confirm that Sgr is like other dSphs in having
no significant rotation ($v_{rot}/\sigma(v) \gtrsim 1 $) signature along any axis (e.g., Figures~\ref{fig:angle}, \S 4).

\subsection{Consistency with the Tidal Stirring Model}

In a companion study by {\L}okas et al. (2010a), $N$-body simulations were generated to 
explain both the elongated shape and velocity characteristics (as found in the present study) 
of the Sgr core.  Of critical significance to the model described there was that the present study
found no significant rotation curve across the face of the dSph, which supports the notion that
Sgr is bar-like with the major axis almost perpendicular to the line of sight, not a disk seen
almost edge-on, which would exhibit a significant rotational signal.  The bulk of the small observed
rotation is simply due to projection effects, and the insignificant intrinsic rotation of the system
points toward a prolate shape for the galaxy, a configuration supported by radially-orbiting stars.
A disk-like configuration was recently employed by Pe\~narrubia et al. (2010) as an intriguing
way to produce a bifurcated leading arm (as a means to explain the two Sgr streams seen in the
Sloan Digital Sky Survey data --- Belokurov et al. 2006; Yanny et al. 2009), but the 20 km s$^{-1}$
 level of rotation expected to be seen today in such a model is clearly precluded by 
 the present observations, a point also made by Pe\~narrubia et al. (2011) from observations they
 made to test their model (over a smaller area than covered here).
 
 On the other hand, as shown in {\L}okas et al. (2010a), the data presented here are consistent
 with the model of tidal stirring (Mayer et al. 2001; Klimentowski et al. 2007, 2009; 
 Kazantzidis et al. 2011) proposed there to account for the shape and dynamics of the 
 Sgr core.  According to the proposed scenario, dwarf galaxies like Sgr did indeed start out
 as disky dwarfs systems (embedded in extended dark matter halos), but in the presence of the
 tidal field of a larger host galaxy like the MW, these systems are transformed into 
 spheroids via formation and subsequent shortening of a bar, and through this process the stellar motions
 are altered from ordered to random.  Depending on what phase of this dynamical transformation
 the dwarf galaxy is viewed, it can be highly elongated (e.g., just after bar formation)
 to very spherical (e.g., much further evolved).  The best fitting model for the Sgr case explored
 by {\L}okas et al. places the presently observed phase just beyond the second perigalacticon, 
 intermediate between the phase of bar formation at first perigalacticon (a phase where Sgr
 would 
 have tidal arms too short) and the third perigalacticon (a phase 
 where Sgr would be spherical).  
 
 This is the first model to explain both the very large ellipticity
 of Sgr as well as the observed kinematics of its stars, but presupposes that the current Sgr orbit
 is a recent product of dynamical friction from a larger orbit having an apocenter exceeding 100 kpc on which
 Sgr was deposited after cosmological infall into the MW halo.  This orbital evolution
 could occur if the initial mass of the system was large --- similar to that of the Large Magellanic 
 Cloud, a system itself that has probably only recently been accreted (e.g., Besla et al. 2007) and that
 presently contains a bar, a morphology consistent with the early phases of the proposed evolutionary 
 scenario.  Coincident with the orbital erosion from dynamical friction, the Sgr system must have shed a significant 
 amount of mass, mostly dark matter, but also probably baryons, to bring it to its presently smaller size; a large,
 LMC-mass progenitor would be required to achieve a significant amount of orbital erosion (Colpi et al.
 1999; Jiang \& Binney 2000; Taffoni et al. 2003).  That the LMC may be an apt model for the Sgr progenitor
 is supported by the similarity in extended star formation histories (e.g., Siegel et al. 2007; Harris \& Zaritsky 2009)
 and detailed chemical evolution (Chou et al. 2010).
 Indeed, as we reiterate from {\L}okas et al. (2010a), 
 the primary difference in the present appearance of the LMC and Sgr may well have to do with timing ---
 i.e., the phase of dynamical evolution dictated by the relative orbital sizes and times they have been bound to 
 the MW.\footnote{For example, Sgr differs from the LMC in being
 presently devoid of gas, but the former clearly had gas as recently as $\sim$0.75 Gyr ago to create its youngest
 stellar population (Siegel et al. 2007).  That last star formation episode may well have depleted Sgr's
 gas reservoir, and/or other processes --- such as ram pressure stripping or supernovae blowout ---
 may have contributed.  All three processes --- (1) gas depletion, (2) ram pressure stripping, and 
 (3) supernovae blowout ---
 are conceivably accelerated in the case of Sgr because of (1) more frequent and intense tidal shocking 
 to produce starbursts at its smaller orbital radius, (2) the higher density of the ram pressure medium at this 
 orbital radius, or (3) the weakened binding energy for the more dynamically evolved, 
 tidally diminished Sgr core.}

\subsection{Is the Disrupting Sagittarius Galaxy a Dwarf Spheroidal Exception or Rosetta Stone?}

In the previous section we made a connection of the Sgr system to dwarf irregular/dwarf spiral galaxies 
via dynamical evolution models with tidal stirring and through observed similarities in star formation
and enrichment history.  But how does Sgr fit within the context of the dwarf spheroidal galaxies with
which it is normally morphologically classified, and what does it potentially tell us about the origin and 
evolution of these systems?

More than seventy years after their discovery, there still remains no 
universally agreed upon, complete picture of the nature of dwarf spheroidal (dSph) 
galaxies.  Hodge (1964a, 1961b, 1962, 1966) and Hodge \& Michie (1969) first invoked tidal 
effects to explain the diffuse structure of at least some dSphs.  
On the other hand, ever since the work of Aaronson (1983) to derive the first 
radial velocity dispersion ($\sigma_{v}$) of a dSph, the notion that they are 
laden with dark matter (DM) has emerged as the standard paradigm to explain
the observed structure and dynamics of these systems. 

In the last decade, enormous efforts have been plied to collecting data on dSphs
to their greatest radial extent.
Evidence for very distended structure 
has been reported for the classical (i.e., the most luminous, not including the
``ultrafaints") MW dSphs (e.g., Majewski et al.\ 2000; Martinez-Delgado et al.\ 2001; 
Palma et al.\ 2003; Westfall et al.\ 2006; Mu\~noz et al.\ 2006; Sohn et
al. 2006; some of these data are collected in Figure~\ref{fig:all_dense}).  
These observed morphologies might be used to infer either that dSphs are
being tidally disrupted, with the stripped stars accounting for the extended structures 
(the general view of the above cited sources; see also {\L}okas et al. 2008), 
or that the dSphs are {\it very} large, fully internally-bound, massive structures
(e.g., Gilmore et al. 2007; Wu 2007).\footnote{In a study of the Sculptor dSph,
Coleman et al. (2005a) suggest yet another interpretation of the extended
structure as one relating to variations in the distribution of stellar populations; 
nevertheless these authors do not rule out the possibility that a small extratidal 
component may exist in this system.  Meanwhile, Coleman et al. (2005b) 
identify an apparent excess of stars around the Fornax system as due to the
remains of another, smaller satellite that previously merged with Fornax.}

\begin{figure} \epsscale{1.21}
\plotone{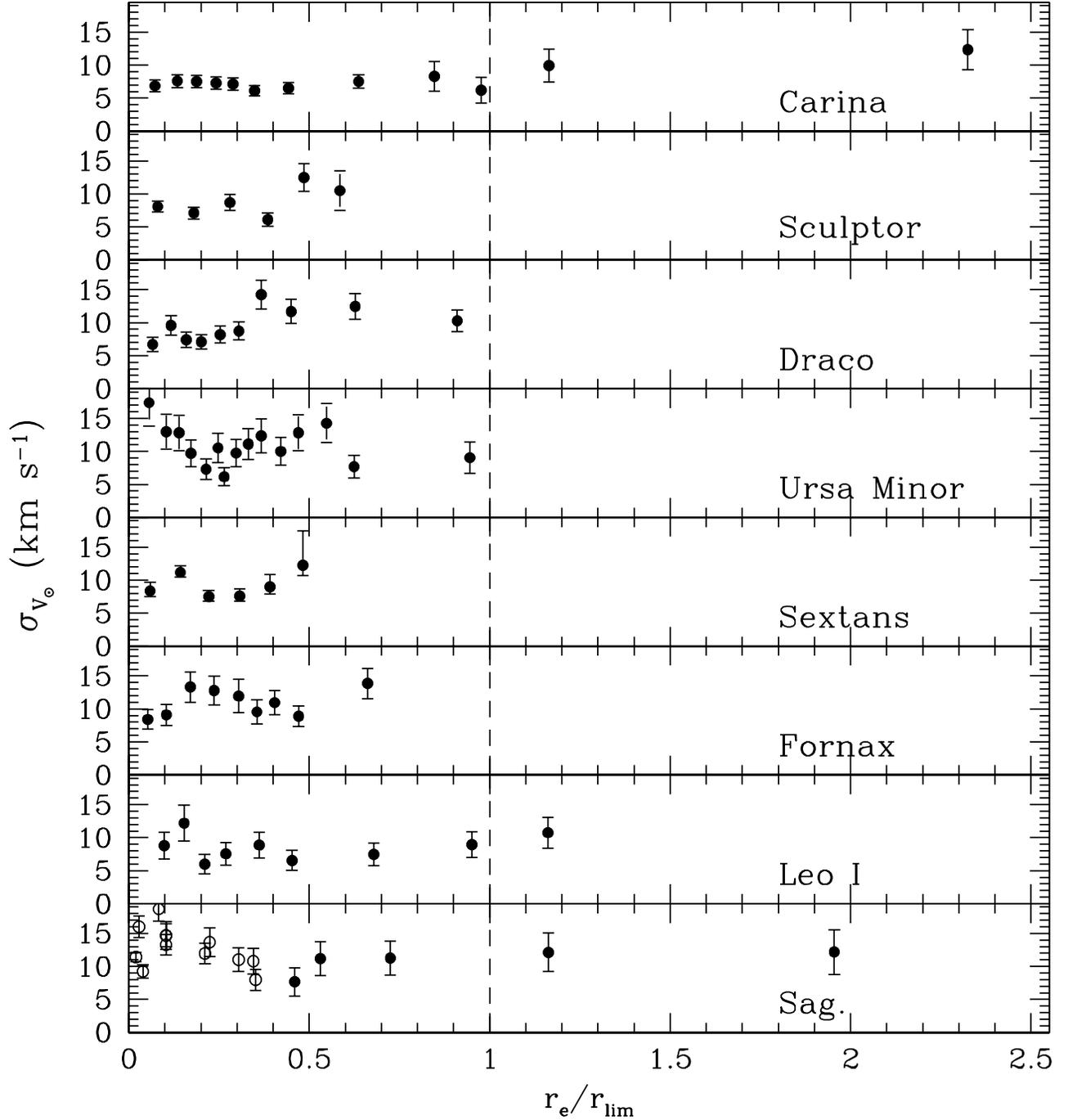}
\caption{\label{fig:all_disp}
Observed trend of velocity dispersion profiles versus projected radius for 
Milky Way dSphs. The vertical line designates the King limiting radii for the
different objects, to which the abscissa has been scaled.
Similar profiles have been reported by other authors (e.g., \citealt{walker07}).
REFERENCES:
Carina (Mu{\~n}oz et al.\ 2006), Sculptor (Westfall et al.\ 2006),
Draco and Ursa Minor (Mu{\~n}oz et al.\ 2005), Sextans (Walker et
al. 2009), Leo I (Sohn
et al.\ 2007), Sagittarius (This paper, Majewski et al. 2004)}
\end{figure}

\begin{figure} \epsscale{1.21}
\plotone{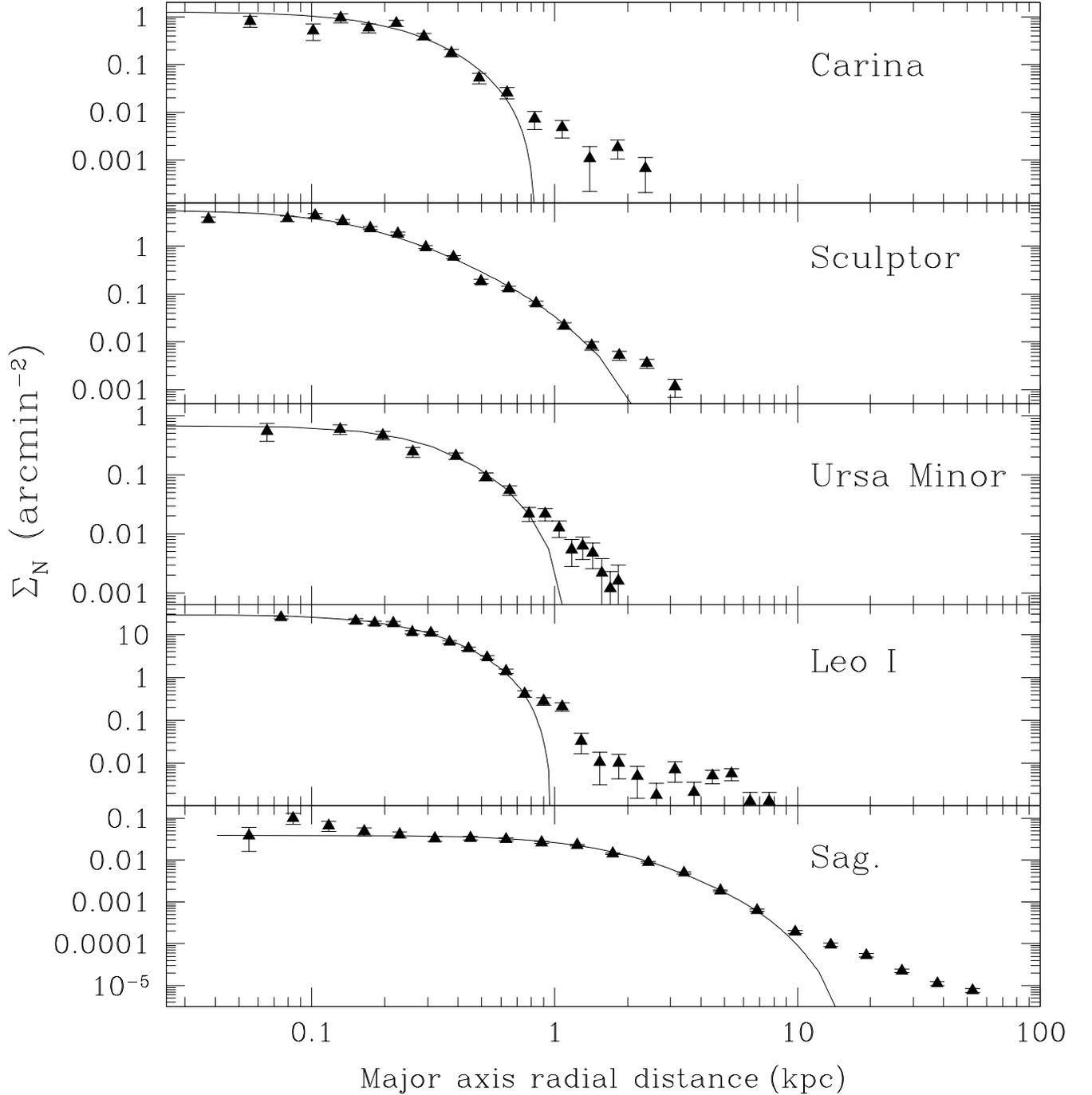}
\caption{\label{fig:all_dense} 
Radial number density profiles of associated red-giant-branch candidates (minus the mean field
giant background) for five dSphs. The solid lines show King profiles fit to the
central regions. Error bars represent Poisson errors. REFERENCES:
Carina (Mu{\~n}oz et al.\ 2006), Sculptor (Westfall et al.\ 2006), Ursa Minor (Palma et al.\ 2003), Leo I (Sohn
et al.\ 2007), Sagittarius (Majewski et al.\ 2003)}
\end{figure}

The latter view has often been bolstered
by the advent of spectroscopic studies 
that have collected hundreds, and in some cases thousands, of individual radial 
velocities over much of the luminous extent of dSphs.  These new data sets have 
revealed flat or just slowly rising/declining velocity dispersion profiles all 
the way to the limits of where data exist (e.g., Westfall et al.\ 2006; Sohn et al.\ 2006; 
Mu\~noz et al.\ 2006; Walker et al.\ 2007 --- some of these data are summarized in 
Fig.~\ref{fig:all_disp}), 
a behavior that might be reasonably well 
explained by the assumption that dSphs live inside much larger DM halos 
in dynamical equilibrium (e.g., Kleyna et al. 2002;
{\L}okas 2002; {\L}okas et al. 2005; Maschenko et al. 2005; Gilmore et al. 2007;
Koch et al. 2007; Wu 2007; Walker et al. 2007)
and resulting in  dramatically rising mass-to-light ratios with radius and large
inferred total satellite masses.

However, as argued by Mu\~noz et al. (2005, 2006), 
the extent of some of the mapped dSphs, as verified through RV-membership
studies of extreme outlying stars in the luminosity profiles,
imply very large linear dimensions, mass-to-light ratios, 
and total masses if all of the identified outer stellar members are assumed to be bound; 
for example, in the case of Carina, Mu\~noz et al. obtain $M/L>16,000$, 
$M_{dSph}=7.2\times10^{9}$ and $R_{tidal}\sim4$\,kpc to keep the outermost RV-member
star bound.
While there is currently debate about whether $M/L$ ratios this large make sense for 
the classical MW dSphs --- e.g., whether they sit in the largest DM subhalos seen
in simulations or in systematically smaller halos (Ferrero et al. 2011; Boylan-Kolchin
et al. 2011) --- the above-implied dimensions for Carina rival those of the Magellanic Clouds
and are not likely to hold universally for all classical MW dSphs based on expected
subhalo mass spectra (e.g., Moore et al. 1999).
It seems even less likely that these dimensions would hold, just by chance, 
for the first few examples of
dSphs that have received the most radially extensive observational attention.

Alternatively, the observed flat velocity dispersion profiles up to and beyond 
the observed King limiting radii, $r_{lim}$, of these systems might also be taken as
further proof for a more universal tendency for the classical MW dSphs, while containing
DM to be sure, to be
experiencing tidal disruption of their luminous parts as well.  It is now well established that the MW halo 
contains extensive stellar substructure from hierarchically merged ``subhalos'', and dSph
galaxies are prime candidates for the progenitors of these substructures.
Moreover, Sohn et al.\ (2006);  Mu\~noz et al.\ (2008); {\L}okas et al. (2008, 2010b), among others, 
have successfully produced $N$-body simulations of DM-filled, tidally disrupting, mass-follows-light 
satellites that well match the 
observed properties of the Carina and Leo I dSphs --- two 
of the three dSph systems (apart from Sgr) with the most extensive radial coverage
in their velocity mappings.  Similar models are also seen to work for Fornax,
Sculptor and Sextans ({\L}okas 2009).

Here we lend further support to this morphological and evolutionary
paradigm from a decidedly phenomenological point of view.  
Figure ~\ref{fig:all_dense}, a synthesis of previous structural studies (see references above),
clearly demonstrates the similarity of the Sgr {\it morphology} to that of other dSphs --- 
i.e., an inner King profile with, at the largest radii, an extended, slower declining 
``break\footnote{That is, ``breaking away'' from the King profile.} population'' --- whereas
the present study demonstrates quite vividly that the 
Sgr dwarf galaxy, {\it a DM-dominated system that nonetheless is 
demonstrably undergoing tidal disruption}, also has 
observed kinematics strongly resembling those of the other classical MW dSph galaxies.
Of course, the lack of any significant rotation in Sgr is commensurate with the kinematics 
of other dSphs.  But, in addition, as
Figure~\ref{fig:all_disp} demonstrates (data from Mu\~noz et al 2005, 2006; Walker et al.\ 2007; and
Sohn et al.\ 2006), the velocity dispersion profile for Sgr as a function of radius for the data 
presented in this study also strongly imitates those for other MW dSphs. 
Just like the velocity dispersion profiles of the other
classical MW dSphs, that of Sgr remains more or less flat 
to large radii, 
including that part of the
system where it is clearly dominated by tidal debris.

As may be seen from the compilation of properties of the classical MW dSph
systems in Table~\ref{Tab:dwarfs}, Sgr is at one extreme within this group in terms
of its distance, luminosity, linear size, ellipticity and derived mass, which might 
merit it being considered ``exceptional'' among the group.  
However, while Sgr is the most
luminous and most massive dSph, it is not significantly brighter or more massive 
than Fornax, and its central velocity dispersion is similar to those of Sculptor, Draco, Ursa Minor,
and Leo I and less than that of Fornax.  Sgr also has an extended star formation history, like Fornax, 
Carina, Sculptor, Sextans, Leo I and Leo II.
Meanwhile, we would argue that Sgr's large ellipticity and half-light radius may well be a 
function of its presently small Galactocentric distance: As discussed above (\S 6.2) the ellipticity
may be due to the recent infall of the system to a small orbit that induced tidal stirring, whereas 
the large present size may be due to Sgr recently experiencing catastrophic tidal disruption
and mass loss, as argued by Chou et al. (2007).

Indeed, guided by the remarkable morphological and dynamical similarity of Sgr to the other MW
dSph systems as well as its compatibility with the tidal stirring model, we conclude 
that Sgr may sometimes
appear to be a ``dSph outlier'' simply because it is presently in a
more flashy, but intermediate phase of a more universal evolutionary sequence 
from a disky, star-forming, ``LMC-like" state to a more spherical, staid, dwarf spheroidal state
long past active star formation, like the Draco system.  As shown by {\L}okas et al. (2010a, 2011),
the exact timescale for a dSph to follow
this evolution is a function of its mass and orbital size, with the latter, of course, a function
of the former through dynamical friction and the zero-age of the evolution set 
by the epoch of initial infall into the MW.  In the case of Sgr, its present relative 
``flashiness'' is driven by its currently small orbit (which amplifies the tidal
stirring effect), proximity to perigalacticon (the most dramatic phase of both tidal stirring 
and mass loss), and its
having recently formed significant stellar tidal tails (again, a function of the stronger 
tidal force experienced on the smaller orbit).  If star formation history
and morphology 
present ersatz timestamps of this ``universal'' 
evolutionary sequence (admittedly crude timestamps, because of the vagaries of 
viewing perspective, orbital shape, stochasticity of star formation histories, and
initial conditions), then perhaps the less bar-like ({\L}okas et al. 2012) and recently star forming
Fornax system may be another system in a somewhat earlier phase
of transformation,
while Carina and Leo I may be in a somewhat later phase. 
This proposed paradigm of a universal evolutionary
track for tidally-stirred, infalling galaxies is explored in more detail, as a function of more
parameters and including all Local Group galaxies in {\L}okas et al. (2011).

Our primary point here is that
because Sgr offers a clear example {\it already found in nature} of a disrupting dwarf galaxy
system containing DM that shares so many properties 
--- e.g., density profile, lack of rotation, $\sigma_v$ profile ---
 with other classical, DM-filled MW dSphs that
it is worth considering that the paradigms of not only tidal influence (e.g., stirring), but tidal 
{\it disruption}, may apply more ubiquitously 
to these other (classical) dSphs and that perhaps these other objects are also suffering varying 
degrees of tidal disruption, along the Sgr paradigm. 
In fact, as found by {\L}okas et al. (2010a), morphological transformation and mass loss
take place together.
Thus, we contend that tidal disruption, as in the case of Sgr, remains
a viable explanation for the observed kinematics and structure at large radii 
of these other dSphs (as already suggested by e.g., Kuhn et al. 1996; 
Majewski et al. 2000; 
Palma et al.\ 2003; Westfall et al.\ 2006; Mu\~noz et al.\ 2006; Sohn et al. 2007; Mateo et al. 2008; 
{\L}okas et al. 2008
and modeled as DM-dominated but tidally influenced systems 
by, e.g., Sohn et al. 2007; Mu\~noz et al. 2006, 2008;
{\L}okas et al. 2008).  Indeed, with Sgr a widely-accepted representative of the 
tidal debris model, but with good cases in hand
also for the Carina and Leo I  systems (modeled well by even simple mass-follows-light DM models),
one may well question the need to have a second explanation --- i.e., large, extended
DM halos to keep the observed extended stellar distributions bound and
dynamically hot --- to explain the outer structure of the other classical MW
dSphs.  In the least, if there are 
two mechanisms responsible for the outer structure and dynamics of dSphs, our results
here show that dynamical studies are hard pressed to discriminate them.  

Moreover, our analysis of the kinematics of the Sgr system demonstrate clearly that 
flat velocity dispersion profiles do {\it not} prove the existence of extended DM halos
around dSphs.  On the other hand, tidal tails, seen morphologically and not just inferred
from kinematics, prove when extended DM halos are not present.  Thus, one 
clear means by which we might hope to distinguish between structural/dynamical models
 --- i.e., extended DM halos versus unbound tidal debris ---  for specific dSphs
is to build more extensive and refined morphological maps capable of detecting
(subtle) tidal tails.  A more sensitive and systematic search for evidence of tidal tails around the 
Galactic dSph population would help build tighter empirical constraints on their true masses
and help answer the questions of whether dSphs typically live
in truly huge ($10^{10-11}$ M$_{\odot}$) DM subhalos or smaller ones and 
what this implies for the efficiency of
galaxy formation on these scales (Ferrero et al. 2011; Boylan-Kolchin et al. 2011).

In conclusion, in contrast to the way Sgr is often portrayed in 
studies of MW dSphs (see Section 1), we argue that it may not be an
anomaly, but rather a Rosetta Stone of dSph galaxies and their 
evolution in a MW-like environment.  Perhaps Sgr's only overriding exceptional quality 
is that it is the MW dSph presently in the most dramatic stages of a universal evolutionary
path that includes tidal harassment and tidal 
disruption.

\acknowledgments
We acknowledge funding by NSF grant AST-0307851, NASA/JPL contract
1228235, the David and Lucile Packard Foundation, and the F.H. Levinson 
Fund of the Peninsula Community Foundation.  
PMF was supported by 
an NSF Astronomy and Astrophysics Postdoctoral Fellowship under award AST-0602221, 
the NASA Graduate Student Researchers Program, a University of Virginia 
Faculty Senate Dissertation-Year Fellowship.
RRM acknowledges support by
the GEMINI-CONICYT Fund, allocated
to the project No. 32080010 and from CONICYT through projects FONDAP N15010003 and
BASAL PFB-06.   EL{\L} is supported by the Polish National Science Centre grant no. N N203 580940.


\clearpage



\begin{deluxetable}{lrrrrrrcc}
\tabletypesize{\scriptsize}
\tablewidth{0pt}
\tablecaption{Field Pointings in Galactic and Sgr Core Coordinates \label{FieldCent}}
\tablehead{ Field & \multicolumn{1}{c}{$\Lambda_{\odot}$\tablenotemark{a}}&
  \multicolumn{1}{c}{$B_{\odot}$\tablenotemark{a}} &  \multicolumn{1}{c}{$l$} &
  \multicolumn{1}{c}{$b$}  &  \multicolumn{1}{c}{$\alpha_{(2000)}$} &
  \multicolumn{1}{c}{$\delta_{(2000)}$} &  \multicolumn{1}{c}{$\left<
      E(B-V) \right>_{Phot}$\tablenotemark{b}} &  \multicolumn{1}{c}{$\left< E(B-V) \right>_{Spec}$\tablenotemark{c}}}
\startdata
  Major$+$00  &   0.115  &   1.615  &   5.51  & $-$14.21 & 18:55:24.29  & $-$30:36:54.0  & 0.151 & 0.154   \\[0.6ex]
  Major$-$04  & 355.992  &   1.221  &   4.92  & $-$10.11 & 18:37:03.48  & $-$29:26:30.7  & 0.201 & 0.197  \\
  Major$-$02  & 358.148  &   1.454  &   5.20  & $-$12.26 & 18:46:34.41  & $-$30:06:03.5  & 0.200 & 0.170   \\
  Major$+$02  &   1.937  &   1.655  &   5.91  & $-$15.99 & 19:03:44.36  & $-$30:57:06.0  & 0.175 & 0.121   \\
  Major$+$04  &   3.873  &   1.777  &   6.26  & $-$17.90 & 19:12:36.40  & $-$31:21:11.1  & 0.100 & 0.101   \\
  Major$+$06  &   5.982  &   2.082  &   6.47  & $-$20.02 & 19:22:15.04  & $-$31:55:09.0  & 0.122 & 0.119   \\
  Major$+$08  &   7.969  &   2.464  &   6.57  & $-$22.04 & 19:31:23.80  & $-$32:30:29.5  & 0.089 & 0.088   \\
  Major$+$10  &   9.992  &   2.658  &   6.88  & $-$24.05 & 19:40:53.36  & $-$32:52:20.0  & 0.156 & 0.159   \\
  Major$+$12  &  12.039  &   3.139  &   6.89  & $-$26.15 & 19:50:29.47  & $-$33:29:01.1  & 0.175 & 0.176   \\[0.6ex]

  Minor$-$03  & 359.721  &   4.448  &   2.57  & $-$14.47 & 18:51:13.29  & $-$33:20:03.2  & 0.091 & 0.089   \\
  Minor$-$02  & 359.723  &   3.276  &   3.75  & $-$14.21 & 18:52:14.06  & $-$32:10:52.7  & 0.091 & 0.125   \\
  Minor$-$01  & 359.994  &   2.473  &   4.62  & $-$14.29 & 18:54:09.13  & $-$31:26:19.4  & 0.154 & 0.155   \\
  Minor$+$01  &   0.090  &   0.571  &   6.55  & $-$13.94 & 18:56:06.96  & $-$29:34:55.6  & 0.154 & 0.162   \\
  Minor$+$02  &   0.248  &$-$0.610  &   7.77  & $-$13.81 & 18:57:44.32  & $-$28:26:40.7  & 0.202 & 0.206   \\
  Minor$+$03  &   0.181  &$-$1.368  &   8.51  & $-$13.56 & 18:58:00.68  & $-$27:41:08.9  & 0.202 & 0.219   \\
  Minor$+$05  &   0.430  &$-$3.494  &  10.69  & $-$13.27 & 19:00:39.99  & $-$25:37:45.8  & 0.224 & 0.204   \\[0.6ex]

     SE$+$04  &   2.541  &   4.394  &   3.27  & $-$17.20 & 19:04:34.06  & $-$33:45:02.3  & 0.101 & 0.100   \\
     SE$+$02  &   1.206  &   2.916  &   4.46  & $-$15.57 & 18:59:24.82  & $-$32:04:48.7  & 0.121 & 0.118   \\
     NW$+$02  & 358.860  &$-$0.122  &   6.94  & $-$12.58 & 18:51:07.53  & $-$28:41:03.5  & 0.177 & 0.173   \\
     NW$+$04  & 357.038  &$-$1.359  &   7.81  & $-$10.87 & 18:45:41.78  & $-$27:11:49.2  & 0.148 & 0.265   \\
     SW$+$04  & 357.038  &   4.128  &   2.28  & $-$11.79 & 18:39:00.43  & $-$32:30:07.8  & 0.124 & 0.121   \\
     SW$+$02  & 358.607  &   2.686  &   4.08  & $-$12.99 & 18:47:35.03  & $-$31:23:49.9  & 0.123 & 0.153   \\
     NE$+$02  &   1.602  &$-$0.003  &   7.50  & $-$15.27 & 19:03:23.59  & $-$29:15:40.7  & 0.157 & 0.118   \\
    ESE$+$07  &   5.341  &   4.836  &   3.46  & $-$20.02 & 19:17:37.87  & $-$34:34:38.3  & 0.095 & 0.095   \\
\enddata
\tablenotetext{a}{The Sgr coordinate system is described in
  \citet{majew03} and is defined by virtue of the orientation
of the extended tidal stream. However we note that (as shown in that same reference; see their Fig.\ 7) 
the main body of Sgr is canted by about $6^{\circ}$ with respect to the mean orientation of
the stream, so that our major and minor axis positions do not directly
line up with the ($\Lambda_{\odot}$, $B_{\odot}$) system.} 
\tablenotetext{b}{$\left< E(B-V) \right>_{Phot}$ measured from
  \citet{schlegel98} for all
  2MASS stars in the selection box for the field.}
\tablenotetext{c}{$\left< E(B-V) \right>_{Spec}$ measured from \citet{schlegel98} for the
  spectroscopically measured Sgr members stars (see Table \ref{Tab:RVstars}).}
\end{deluxetable}

\begin{deluxetable}{lcrrrrrrrrrcc}
\tabletypesize{\tiny}
\tablewidth{0pt}
\tablecaption{\label{Tab:RVstars} Sagittarius Stellar Velocity Data}
\tablehead{
\colhead{Field}&\colhead{NAME}&     \colhead{$K_{s,o}$}&    \colhead{$\!\!\!(J$-$K_s)_o\!\!\!$}&   \colhead{$E_{B-V}$\tablenotemark{a}}&   \colhead{$l$}&      \colhead{$b$}&       \colhead{$\Lambda_{\odot}$}& \colhead{$V_{hel}$} &  \colhead{$V_{GSR}$}&  \colhead{$\epsilon_{V}$} & \colhead{$TDR$} & \colhead{Member?} \\
\colhead{} &\colhead{} & \colhead{}    & \colhead{}   & \colhead{}  &   \colhead{(deg)}&      \colhead{(deg)}&     \colhead{(deg)}& \colhead{$\!\!\!$(km s$^{-1}$)$\!\!\!$} &  \colhead{$\!\!\!$(km s$^{-1}$)$\!\!\!$} &  \colhead{$\!\!\!$(km s$^{-1}$)$\!\!\!$} & \colhead{} & \colhead{}
 }
\startdata 
Major$+$00 & 18541047$-$3026161 & 12.47 & 0.83 & 0.16  & 5.5688 & $-$13.8964 & 359.824  & $+$139.1  & $+$168.2 & 2.2 & 15.91 & Y  \\
Major$+$00 & 18541810$-$3029314 & 12.11 & 0.98 & 0.15  & 5.5288 & $-$13.9432 & 359.860  & $+$146.8  & $+$175.8 & 1.8 & 16.64 & Y  \\
Major$+$00 & 18543205$-$3047488 & 12.03 & 0.82 & 0.15  & 5.2612 & $-$14.1105 & 359.962  & $-$14.7  &  $+$13.2 & 2.6 & 11.15 & N  \\
Major$+$00 & 18550269$-$3045523 & 12.43 & 0.94 & 0.16  & 5.3367 & $-$14.1983 &   0.065  & $+$154.0  & $+$182.2 & 1.8 & 20.41 & Y  \\
Major$+$00 & 18550257$-$3025397 & 12.60 & 0.75 & 0.15  & 5.6553 & $-$14.0640 &   0.006  & $+$144.4  & $+$173.9 & 2.0 & 16.13 & Y  \\
Major$+$00 & 18542551$-$3039455 & 12.53 & 0.88 & 0.15  & 5.3785 & $-$14.0356 & 359.916  & $+$140.7  & $+$169.1 & 2.3 & 18.08 & Y  \\
Major$+$00 & 18545211$-$3024229 & 12.67 & 0.79 & 0.15  & 5.6600 & $-$14.0211 & 359.966  & $+$139.2  & $+$168.7 & 2.2 & 12.96 & Y  \\
Major$+$00 & 18542283$-$3051089 & 12.25 & 0.85 & 0.15  & 5.1951 & $-$14.1023 & 359.940  &  $+$14.5  &  $+$42.2 & 2.7 & 14.92 & N  \\
Major$+$00 & 18553060$-$3047381 & 11.89 & 1.02 & 0.17  & 5.3498 & $-$14.3017 &   0.168  & $+$143.2  & $+$171.5 & 1.9 & 24.95 & Y  \\
Major$+$00 & 18552942$-$3034507 & 10.58 & 1.12 & 0.15  & 5.5499 & $-$14.2133 &   0.128  & $+$134.5  & $+$163.5 & 1.6 & 34.25 & Y  \\
Major$+$00 & 18553188$-$3041275 & 13.00 & 0.69 & 0.16  & 5.4491 & $-$14.2651 &   0.155  &  $+$40.6  &  $+$69.2 & 1.7 & 10.74 & N  \\
\multicolumn{13}{c}{.........}\\
\enddata 
\tablenotetext{a}{$E(B-V)$ measured from \citet{schlegel98}.} 
\tablenotetext{b}{Table 2 is published in its entirety in the
  electronic edition of the Astrophysical Journal.  A portion is shown here for guidance regarding its form and content.}
\end{deluxetable}

\begin{deluxetable}{lcrrrrrrr}
\tabletypesize{\scriptsize}
\tablewidth{0pt}
\tablecaption{\label{Tab:RVs} Field Kinematic Parameters}
\tablehead{Field & $\langle V_{GSR} \rangle$ (km s$^{-1}$)
&  $\sigma_{V_{GSR}}$ (km s$^{-1}$) & $R$ (') & $R_e$ (') & RV stars}
\startdata
Major$+$00     &  170.14 $\pm$ 1.00  &    9.85 $\pm$ 0.73 &   10.7 &   23.1 &   102 \\[1ex]
Major$-$04     &  175.02 $\pm$ 2.09  &   15.60 $\pm$ 1.49 &  239.7 &  239.8 &    57  \\
Major$-$02     &  172.91 $\pm$ 1.38  &   12.66 $\pm$ 0.98 &  110.9 &  112.0 &    86  \\
Major$+$02     &  170.11 $\pm$ 1.01  &   13.39 $\pm$ 0.73 &  116.9 &  117.0 &   180   \\
Major$+$04     &  167.93 $\pm$ 1.38  &   12.25 $\pm$ 1.00 &  234.0 &  234.3 &    82  \\
Major$+$06     &  164.20 $\pm$ 1.55  &   11.44 $\pm$ 1.13 &  363.2 &  363.3 &    57  \\
Major$+$08     &  165.19 $\pm$ 1.62  &   10.45 $\pm$ 1.18 &  486.5 &  486.6 &    44  \\
Major$+$10     &  157.56 $\pm$ 2.97  &   14.34 $\pm$ 2.13 &  610.9 &  611.1 &    24  \\
Major$+$12     &  152.15 $\pm$ 3.67  &   13.50 $\pm$ 2.62 &  740.2 &  740.2 &    14  \\[1ex]
Minor$-$03     &  167.40 $\pm$ 3.19  &   16.40 $\pm$ 2.27 &  179.2 &  511.7 &   27   \\
Minor$-$02     &  173.58 $\pm$ 2.23  &   16.52 $\pm$ 1.58 &  109.2 &  311.1 &   56  \\
Minor$-$01     &  173.33 $\pm$ 1.39  &   12.90 $\pm$ 1.00 &   59.7 &  170.1 &   89  \\
Minor$+$01     &  172.73 $\pm$ 2.06  &   15.15 $\pm$ 1.48 &   54.7 &  156.2 &   56   \\
Minor$+$02     &  167.46 $\pm$ 2.29  &   15.75 $\pm$ 1.64 &  126.1 &  360.0 &   48   \\
Minor$+$03     &  156.28 $\pm$ 3.99  &   26.91 $\pm$ 2.83 &  171.0 &  488.5 &   46   \\
Minor$+$05     &  154.54 $\pm$ 6.13  &   18.22 $\pm$ 4.36 &  299.0 &  854.4 &    9   \\[1ex]
SE$+$04        &  163.94 $\pm$ 2.51  &   14.44 $\pm$ 1.80 &  233.1 &  489.2 &   34   \\
SE$+$02        &  171.20 $\pm$ 1.38  &   12.78 $\pm$ 0.99 &  112.8 &  242.1 &   88   \\
NW$+$02        &  170.46 $\pm$ 1.66  &   12.90 $\pm$ 1.19 &  117.7 &  270.4 &   62   \\
NW$+$04        &  175.95 $\pm$ 5.24  &   20.86 $\pm$ 3.75 &  229.7 &  486.1 &   16   \\[1ex]
 
SW$+$04        &  167.78 $\pm$ 2.62  &    8.19 $\pm$ 1.96 &  240.2 &  513.9 &   11   \\
SW$+$02        &  174.51 $\pm$ 1.70  &   12.77 $\pm$ 1.24 &  111.0 &  236.7 &   60   \\
NE$+$02        &  168.50 $\pm$ 2.53  &   15.03 $\pm$ 1.83 &  130.4 &  289.5 &   37   \\[1ex]
ESE$+$07       &  166.29 $\pm$ 2.70  &   13.34 $\pm$ 1.93 &  383.1 &  588.7 &   25  \\[0.5ex]\hline\\[-6pt]
12$^{\circ}$-15$^{\circ}$ &  148.34 $\pm$ 2.79 &  9.05 $\pm$ 2.02 &  864 &   895 &  11  \\
15$^{\circ}$-20$^{\circ}$ &  139.52 $\pm$ 3.95 & 11.67 $\pm$ 2.83 & 1045 &  1073 &   9  \\
20$^{\circ}$-40$^{\circ}$ &  117.64 $\pm$ 4.92 & 16.93 $\pm$ 3.50 & 1675 &  1748 &  12  \\[0.5ex]\hline\\[-6pt]
Ibata f1 AAT   &  155.45 $\pm$ 2.37  &    7.98 $\pm$ 1.60 &  718 &  638 &   15  \\
Ibata f2 AAT   &  165.29 $\pm$ 3.18  &   11.03 $\pm$ 1.81 &  495 &  552 &    9  \\
Ibata f3 AAT   &  168.95 $\pm$ 2.46  &   13.66 $\pm$ 2.19 &  455 &  409 &   17  \\
Ibata f4 AAT   &  172.55 $\pm$ 1.92  &   11.96 $\pm$ 1.58 &  367 &  383 &   26   \\
Ibata f5 AAT   &  172.15 $\pm$ 2.24  &   14.70 $\pm$ 2.12 &   56 &   69 &   30   \\
Ibata f6 AAT   &  172.15 $\pm$ 2.33  &   14.70 $\pm$ 1.76 &   77 &  186 &   30   \\
Ibata f7 AAT   &  169.05 $\pm$ 2.32  &   16.01 $\pm$ 1.62 &   32 &   36 &   44   \\
Ibata f8 AAT   &  167.84 $\pm$ 3.30  &   18.70 $\pm$ 1.80 &      &      &   33   \\[1ex]
Ibata f1 CTIO  &  159.58 $\pm$ 2.06  &   10.80 $\pm$ 1.93 &  718 &  638 &   20   \\
Ibata f5 CTIO  &  171.51 $\pm$ 1.22  &    9.24 $\pm$ 1.04 &   56 &   69 &   48   \\
Ibata f6 CTIO  &  169.53 $\pm$ 2.76  &   13.29 $\pm$ 1.54 &   77 &  186 &   24   \\
Ibata f7 CTIO  &  170.99 $\pm$ 0.80  &   11.41 $\pm$ 0.73 &   32 &   36 &  114  \\[1ex]
Bellazzini (Sgr,N) &  168.03 $\pm$ 0.60  &     9.6 $\pm$ 0.4 &    0   &  0 & 318 \\
\enddata
\end{deluxetable}

\begin{deluxetable}{lccrrrr}
\tabletypesize{\scriptsize}
\tablewidth{0pt}
\tablecaption{Expected Galactic Contamination\label{Contamination}}
\tablehead{  &  \multicolumn{1}{c}{Besan\c{c}on Stars} & \underline{This Study} & \multicolumn{1}{c}{Scale} &
  \multicolumn{1}{l}{Expected} & \underline{This Study} & \multicolumn{1}{l}{Percent}\\\cline{2--2}
  Field & Total & Non-Members & \multicolumn{1}{c}{Factor} &
    \multicolumn{1}{l}{Contaminants} & Members &
    \multicolumn{1}{l}{Contam.\ } }
\startdata
    Major$+$00    &  228    &     18   & 12.67 &  2.4 stars & 102 &   2.4\% \\ 
    Major$-$04    &  573    &     57   & 10.05 &  9.8 stars & 110 &   8.9\% \\ 
    Major$-$02    &  313    &     34   &  9.21 &  4.1 stars &  86 &   4.8\% \\ 
    Major$+$02    &  210    &     34   &  6.18 &  2.6 stars & 182 &   1.4\% \\ 
    Major$+$04    &  181    &     22   &  8.23 &  1.1 stars &  82 &   1.3\% \\ 
    Major$+$06    &  150    &     29   &  5.17 &  2.7 stars &  57 &   4.7\% \\ 
    Major$+$08    &  146    &     42   &  3.48 &  2.3 stars &  45 &   5.1\% \\ 
    Major$+$10    &  120    &     55   &  2.18 &  1.8 stars &  24 &   7.5\% \\ 
    Major$+$12    &  114    &     50   &  2.28 &  0.9 stars &  14 &   6.4\% \\ 
 {\it Minor$-$03}    & {\it 247}    &  {\it   91}   & {\it 2.71} &{\it 25.8 stars} &  {\it 21} & {\it 122.9\%} \\ 
    Minor$-$02    &  230    &     53   &  4.34 &  3.2 stars &  56 &   5.7\% \\ 
    Minor$-$01    &  213    &     26   &  8.19 &  2.8 stars &  89 &   3.1\% \\ 
    Minor$+$01    &  218    &     35   &  6.23 &  4.0 stars &  56 &   7.1\% \\ 
    Minor$+$02    &  264    &     68   &  3.88 &  8.3 stars &  48 &  17.3\% \\ 
    Minor$+$03    &  240    &    104   &  2.31 &  9.1 stars &  49 &  18.6\% \\ 
 {\it Minor$+$05}    & {\it 244}    &   {\it  78 }  & {\it 3.13} & {\it 10.2 stars} & {\it  9} & {\it 113.0\%} \\ 
       SW$+$04    &  354    &     29   & 12.21 &  2.0 stars &  12 &  16.7\% \\  
       SW$+$02    &  256    &     15   & 17.07 &  1.5 stars &  60 &   2.5\% \\ 
       NE$+$02    &  203    &     12   & 16.92 &  1.1 stars &  37 &   2.3\% \\ 
 {\it    NW$+$04}    & {\it 369}    &   {\it  60}   & {\it 6.15} & {\it 10.7 stars} & {\it 17} & {\it 62.9\%} \\ 
       NW$+$02    &  270    &     49   &  5.51 &  8.5 stars &  62 &  13.7\% \\ 
       SE$+$02    &  207    &      25  &  8.28 &  1.9 stars &  88 &   2.2\% \\ 
       SE$+$04    &  198    &      45  &  4.40 &  4.6 stars &  34 &  13.5\% \\ 
      ESE$+$07    &  156    &      66  &  2.36 &  3.8 stars &  25 &  15.2\% \\
\enddata
\end{deluxetable}

\begin{deluxetable}{rrrrrrrrrrrc}
\tabletypesize{\footnotesize}
\tablewidth{0pt}
\tablecaption{Supplementary Radial Velocity Data on M Giants Observed with the DuPont Telescope\label{modspec}}
\tablehead{
\colhead{NAME}&     \colhead{$K_{s,o}$}&
\colhead{$\!\!\!(J$-$K_s)_o\!\!\!$}&   \colhead{$E{B-V}$\tablenotemark{a}}&
\colhead{$l$}&      \colhead{$b$}&       \colhead{$\Lambda_{\odot}$}&
\colhead{$V_{hel}$} &  \colhead{$V_{GSR}$}&  \colhead{$CCP$} &
\colhead{$Q$} &
\colhead{Member?} \\
\colhead{} & \colhead{}    & \colhead{}   & \colhead{}  &
\colhead{(deg)}&      \colhead{(deg)}&     \colhead{(deg)}&
\colhead{$\!\!\!$(km s$^{-1}$)$\!\!\!$} &  \colhead{$\!\!\!$(km
  s$^{-1}$)$\!\!\!$} &  \colhead{} & \colhead{} & \colhead{}
  }

\startdata
$19531855-3448555$& 12.35&  0.97& 0.10&   5.6& $ -27.1$&    12.7&     $  126.1$& $  151.3$& 0.50& 7  &  Y  \\
$19534690-3459146$& 12.08&  0.95& 0.10&   5.5& $ -27.2$&    12.8&     $  124.6$& $  149.3$& 0.45& 6  &  Y  \\
$19541973-3525442$& 11.66&  0.98& 0.10&   5.0& $ -27.4$&    12.9&     $  142.8$& $  165.7$& 0.44& 7  &  Y  \\
$19544480-3431024$& 11.25&  1.05& 0.10&   6.1& $ -27.3$&    13.0&     $  115.7$& $  142.5$& 0.74& 7  &  Y  \\
$19553880-3547483$& 10.96&  1.07& 0.10&   4.7& $ -27.8$&    13.2&     $  135.3$& $  157.0$& 0.71& 7  &  Y  \\
$19571379-3359130$& 10.99&  1.03& 0.10&   6.8& $ -27.6$&    13.5&     $  114.2$& $  143.3$& 0.69& 7  &  Y  \\
$19572127-3626092$& 11.05&  1.03& 0.10&   4.1& $ -28.3$&    13.6&     $  132.3$& $  151.9$& 0.81& 7  &  Y  \\
$19591632-3411107$& 11.65&  1.02& 0.10&   6.7& $ -28.1$&    13.9&     $  -83.1$& $  -54.5$& 0.27& 4  &  N  \\
$20015401-3558348$&  9.76&  1.18& 0.10&   4.9& $ -29.1$&    14.5&     $  130.4$& $  152.2$& 0.79& 7  &  Y  \\
$20041192-3437541$& 10.85&  1.07& 0.10&   6.5& $ -29.2$&    14.9&     $  123.9$& $  151.5$& 0.70& 7  &  Y  \\
$20041677-3508590$& 10.27&  1.14& 0.10&   5.9& $ -29.3$&    15.0&     $  111.3$& $  136.7$& 0.73& 7  &  Y  \\
$20044613-3227255$& 10.68&  1.08& 0.10&   8.9& $ -28.7$&    15.0&     $   94.2$& $  130.3$& 0.75& 7  &  Y  \\
$20050480-3311517$& 10.47&  1.13& 0.10&   8.1& $ -29.0$&    15.1&     $  111.6$& $  144.9$& 0.76& 7  &  Y  \\
$20071791-3223480$& 12.39&  0.93& 0.20&   9.2& $ -29.2$&    15.5&     $   95.2$& $  131.9$& 0.66& 7  &  Y  \\
$20094440-3501218$& 10.90&  1.08& 0.10&   6.4& $ -30.4$&    16.1&     $  105.2$& $  131.7$& 0.83& 7  &  Y  \\
$20104667-3327077$& 10.67&  1.09& 0.10&   8.2& $ -30.2$&    16.3&     $  110.8$& $  143.7$& 0.79& 7  &  Y  \\
$20122492-3310058$& 10.96&  1.06& 0.20&   8.6& $ -30.5$&    16.6&     $  112.5$& $  146.7$& 0.74& 7  &  Y  \\
$20134595-3421279$& 10.96&  1.06& 0.10&   7.4& $ -31.0$&    16.9&     $   88.8$& $  118.4$& 0.84& 7  &  Y  \\
$20183323-3807468$& 11.45&  0.99& 0.10&   3.2& $ -32.8$&    17.9&     $  144.2$& $  154.3$& 0.60& 7  &  Y  \\
$20214447-3353414$& 11.04&  1.04& 0.10&   8.3& $ -32.5$&    18.6&     $   96.3$& $  128.5$& 0.75& 7  &  Y  \\
$20233705-3344500$& 11.36&  1.08& 0.10&   8.6& $ -32.9$&    18.9&     $  122.7$& $  155.6$& 0.68& 7  &  Y  \\
$20313337-3244528$& 11.48&  1.04& 0.10&  10.2& $ -34.3$&    20.6&     $  -29.2$& $    8.1$& 0.82& 7  &  N  \\
$20330508-3416310$& 10.53&  1.13& 0.00&   8.5& $ -34.9$&    20.9&     $  117.0$& $  148.5$& 0.82& 7  &  Y  \\
$20372047-3423140$& 11.44&  1.03& 0.00&   8.5& $ -35.8$&    21.8&     $  104.7$& $  135.9$& 0.65& 7  &  Y  \\
$20384078-3911581$& 11.15&  1.00& 0.00&   2.7& $ -36.8$&    21.8&     $  105.6$& $  117.5$& 0.73& 7  &  Y  \\
$20412659-3613296$& 11.49&  1.00& 0.00&   6.5& $ -36.9$&    22.5&     $  113.6$& $  137.6$& 0.81& 7  &  Y  \\
$20452019-3625220$& 11.16&  0.99& 0.10&   6.4& $ -37.7$&    23.3&     $  106.8$& $  130.1$& 0.79& 7  &  Y  \\
$20485427-3510562$& 10.87&  1.07& 0.10&   8.1& $ -38.3$&    24.1&     $   87.5$& $  115.8$& 0.78& 7  &  Y  \\
$20490681-3547523$& 11.35&  1.01& 0.10&   7.3& $ -38.4$&    24.1&     $   94.9$& $  120.7$& 0.90& 7  &  Y  \\
$20514595-3308115$& 10.59&  1.14& 0.10&  10.8& $ -38.5$&    24.9&     $   74.4$& $  111.0$& 0.81& 7  &  Y  \\
$21083956-3755451$& 10.46&  1.11& 0.00&   5.1& $ -42.5$&    27.8&     $   80.5$& $   97.6$& 0.95& 7  &  Y  \\
$21304974-3440522$& 10.53&  1.11& 0.10&  10.1& $ -46.8$&    32.7&     $   -1.4$& $   27.3$& 0.76& 7  &  N  \\
$21343307-3436437$& 10.12&  1.24& 0.00&  10.3& $ -47.5$&    33.5&     $   79.7$& $  108.6$& 0.74& 7  &  Y  \\
$21545225-3331355$& 10.81&  1.11& 0.00&  12.3& $ -51.7$&    37.8&     $   61.5$& $   92.1$& 0.86& 7  &  Y  \\
$22011809-3316344$& 11.31&  0.98& 0.00&  12.8& $ -53.0$&    39.2&     $   65.6$& $   96.3$& 0.88& 7  &  Y  \\
$22263275-3404084$& 11.56&  1.03& 0.00&  11.3& $ -58.2$&    44.1&     $ -133.2$& $ -110.4$& 0.70& 7  &  N  \\
$22514994-3659556$& 11.06&  1.08& 0.00&   4.0& $ -62.9$&    48.2&     $ -126.0$& $ -120.7$& 0.92& 7  &  N  \\

\enddata
\tablenotetext{a}{$E(B-V)$ measured from \citet{schlegel98}.} 
\end{deluxetable}

\begin{deluxetable}{lrcrrrrr}
\tabletypesize{\scriptsize}
\tablewidth{0pt}
\tablecaption{Properties of the Classical Milky Way dSph Galaxies\label{Tab:dwarfs}}
\tablehead{  
\multicolumn{1}{c}{Name} &
\multicolumn{1}{c}{$Distance$} &
\multicolumn{1}{c}{$L_V$\tablenotemark{b}} &
\multicolumn{1}{c}{$r_{half}$\tablenotemark{b}} &
\multicolumn{1}{c}{$\epsilon$\tablenotemark{c}} &
\multicolumn{1}{c}{$\sigma_{V,central}$}&
\multicolumn{1}{c}{$M(r_{half})$\tablenotemark{b}} \\
\multicolumn{1}{c}{$$} &
\multicolumn{1}{c}{$(kpc)$} &
\multicolumn{1}{c}{$(L_{V,\odot})$} &
\multicolumn{1}{c}{$(pc)$} &
\multicolumn{1}{c}{$$} &
\multicolumn{1}{c}{$(km s^{-1})$} &
\multicolumn{1}{c}{$(M_{\odot})$} 
}
\startdata
Leo II       & 219 & $5.9 \times 10^5$ &  151  &  0.13 & $\phantom{1}6.6 \pm 0.7$ &   $3.8 \times 10^6$  \\
Carina       &  72 & $2.4 \times 10^5$ &  241  &  0.33 & $\phantom{1}6.6 \pm 1.2$ &   $6.1 \times 10^6$  \\
Sextans      &  95 & $4.1 \times 10^5$ &  682  &  0.35 & $\phantom{1}7.7 \pm 1.3$ &   $2.5 \times 10^7$  \\
Draco        &  69 & $2.7 \times 10^5$ &  196  &  0.29 & $\phantom{1}9.1 \pm 1.2$ &   $9.4 \times 10^6$  \\
Leo I        & 254 & $3.4 \times 10^6$ &  246  &  0.21 & $\phantom{1}9.2 \pm 1.4$ &   $1.2 \times 10^7$  \\
Sculptor     &  85 & $1.4 \times 10^6$ &  260  &  0.32 & $\phantom{1}9.2 \pm 1.1$ &   $1.3 \times 10^7$  \\
Ursa Minor   &  93 & $2.0 \times 10^5$ &  280  &  0.56 & $\phantom{1}9.5 \pm 1.2$ &   $1.5 \times 10^7$  \\
Sagittarius  &  29 & $1.7 \times 10^7$ & 1550  &  0.65 & $\phantom{1}9.9 \pm 0.7$ &   $1.2 \times 10^8$  \\
Fornax       & 139 & $1.4 \times 10^7$ &  668  &  0.30 &           $11.7 \pm 0.9$ &   $5.3 \times 10^7$  \\
\enddata
\tablenotetext{a}{References: Distances adopted from Rizzi et
  al. (2007) for Fornax, Bellazzini et al. (2004) for Leo I,
  Pietrzy{\'n}ski et al. (2008) for Sculptor, Siegel \& Majewski (2006) for
  Leo II, Lee et al. (2003) for Sextans, Bellazzini et al. (2002) for
  Draco, Mighell \& Burke (1999) for Ursa Minor and Siegel et al. (2011) for Sagittaurius.}
\tablenotetext{b}{Values taken from Walker et al. (2010), see original
paper for references therein}
\tablenotetext{c}{References: Ellipticity adopted from 
  Irwin \& Hatzidimitriou (1995) for Sextans, Fornax,and Leo II 
  Sohn et al. (2007) for Leo I,
  Westfall et al. (2006) for Sculptor, 
  Odenkirchen et al. (2001) for  Draco,
  Palma et al. (2003) for Ursa Minor, 
  and PAPER I for Sagittaurius.} 
\tablenotetext{d}{References: Central Velocity Dispersion adopted from 
  Walker et al. (2009) for Fornax, and Sextans, 
  Sohn et al. (2007) for Leo I,
  Westfall et al. (2006) for Sculptor, 
  Mateo et al. (2008) for Leo II,
  Odenkirchen et al. (2001) for  Draco,
  Palma et al. (2003) for Ursa Minor, 
  and This paper for Sagittaurius.} 
\end{deluxetable}

\clearpage
\end{document}